\documentclass[fleqn,usenatbib]{mnras}

\usepackage{newtxtext,newtxmath}

\usepackage[T1]{fontenc}

\DeclareRobustCommand{\VAN}[3]{#2}
\let\VANthebibliography\thebibliography
\def\thebibliography{\DeclareRobustCommand{\VAN}[3]{##3}\VANthebibliography}

\usepackage{graphicx}	
\usepackage{amsmath}	
\usepackage{enumitem}

\title[NuSTAR view of 4U 1538-522]{Torque reversal and cyclotron absorption feature in HMXB 4U 1538-522}

\author[P. Sharma, C. Jain, and A. Dutta.]{
Prince Sharma,$^{1}$\thanks{E-mail: princerajsharma31@gmail.com}
Chetana Jain$^{2}$\thanks{E-mail: chetanajain11@gmail.com}
and Anjan Dutta$^{1}$\thanks{E-mail: dutta.anjan33@gmail.com}
\\
$^{1}$Department of Physics and Astrophysics, University of Delhi, Delhi 110007, India\\
$^{2}$Hansraj College, University of Delhi, Delhi 110007, India\\
}

\date{Accepted XXX. Received YYY; in original form ZZZ}

\pubyear{2023}

\begin{document}
\label{firstpage}
\pagerange{\pageref{firstpage}--\pageref{lastpage}}
\maketitle

\begin{abstract}
We present a comprehensive timing and spectral analysis of the HMXB 4U 1538-522 by using the \emph{Nuclear Spectroscopic Telescope Array (NuSTAR)} observatory data. Using three archived observations made between 2019 and 2021, we have detected $\sim $ 526 s coherent pulsations up to 60 keV. We have found an instantaneous spin-down rate of $\dot{P} = 6.6_{-6.0}^{+2.4} \times 10^{-6}$ s s$^{-1}$ during the first observation. The pulse profiles had a double peaked structure consisting of a broad primary peak and an energy dependent, weak secondary peak. We have also analysed the long-term spin-period evolution of 4U 1538-522 from data spanning more than four decades, including the data from \emph{Fermi}/GBM. Based on the recent spin trends, we have found that the third torque reversal in 4U 1538-522 happened around MJD 58800. The source is currently spinning up with $\dot{P} = -1.9(1) \times 10^{-9}$ s s$^{-1}$. We also report a periodic fluctuation in the spin-period of 4U 1538-522. The broad-band persistent spectra can be described with a blackbody component and either powerlaw or Comptonization component along with a Fe K$_{\alpha}$ line at 6.4 keV and a cyclotron absorption feature around 22 keV. We have also found a relatively weak absorption feature around 27 keV in the persistent spectra of 4U 1538-522 in all three observations. We have estimated a magnetic field strength of $1.84_{-0.06}^{+0.04} (1+z) \times 10^{12}$ and $2.33_{-0.24}^{+0.15} (1+z) \times 10^{12}$ G for the two features, respectively.  
\end{abstract}




\section{Introduction}

\begin{table*}
	\caption{Log of \emph{NuSTAR} observations used in this work.}
	\label{tab:obslog}
	\begin{tabular*}{\textwidth}{@{\extracolsep{\fill}} cccc|c|c|c|c}
	\hline
		Observation & Observation ID & Observation Date & MJD & Exposure$^{a}$  & \multicolumn{2}{c|}{Count Rate (c/s)} & Orbital Phase$^{b}$\\
		& & dd-mm-yyyy & (d) & (ks) & FPMA & FPMB &\\ 
		\hline
		1 & 30401025002 & 22-05-2019 & 58605.94  & 36.87 & $11.69 \pm 0.02 $ & $ 10.77 \pm 0.02 $ & 0.486--0.727 \\[0.5ex]
		2 & 30602024002 & 16-02-2021 & 59261.32  & 21.81 & $8.61 \pm 0.02 $ & $ 8.07 \pm 0.02 $ & 0.272--0.409 \\[0.5ex]
		3 & 30602024004 & 22-02-2021 & 59267.03  & 21.56 & $ 5.41 \pm 0.02 $ & $ 5.11 \pm 0.02 $ & 0.803--0.948\\[0.5ex]

	\hline
	\multicolumn{7}{l}{\textit{Notes.} $^{a}$Raw exposure of the data.}\\
    \multicolumn{7}{l}{$^{b}$ Based on the orbital ephemeris of \citet{Hemphill2019}.}\\
	\end{tabular*}
\end{table*}



4U 1538-522 is an eclipsing, wind-fed persistent High-mass X-ray binaries (HMXBs), which was discovered with \emph{Uhuru} \citep{Giacconi1974}. X-ray pulsations from the source were detected at 529 s, independently with \emph{Ariel V} \citep{Davison1977a} and \emph{OSO-8} \citep{Becker1977}. Based on the Doppler modulations of the spin period, \citet{Becker1977} and \citet{Davison1977b} established the binary and eclipsing nature of the source. Initial identification suggested a 14.5 mag B0 supergiant companion \citep{Cowley1977,Crampton1978,Parkes1978} at a distance of $5.5 \pm 1.5$ kpc. Subsequent optical studies of the companion refined the spectral type to a B0Iab star with a source distance of $6.4 \pm 1.0 $ kpc \citep{Reynolds1992}. The latest \emph{Gaia} parallax measurements indicate a source distance of $6.6_{-1.5}^{+2.2}$ kpc \citep{Bailer2018}.

The orbital period of  4U 1538-522 is $\sim$ 3.7 d including a 0.6 d long eclipse phase and source inclination of $67^{\circ} \pm 1^{\circ}$ \citep{Becker1977,Makishima1987,Clark2000,Baykal2006,Mukherjee2006,Falanga2015,Hemphill2019}. Despite the well established orbital parameters the exact nature of the orbit remains inconclusive. Some studies favour a circular orbit \citep{Makishima1987,Corbet1993,Baykal2006} while others are biased in the favour of an eccentricity of 0.17-0.18 \citep{Clark2000,Mukherjee2006}. \citet{Rawls2011} have estimated a neutron star (NS) mass of $1.104 \pm 0.177$ M$_{\sun}$ for a circular orbit and $0.87 \pm 0.07$ M$_{\sun}$ for the elliptical solution of \citet{Clark2000}. In either case, the NS mass is unusually low. 

Early results on the spin-period studies of 4U 1538-522 with \emph{Tenma} and \emph{EXOSAT} data showed a random-walk behaviour and it acted as an evidence in support of a wind-fed pulsar \citep{Makishima1987,Cusumano1989}. During 1976-1988, this random walk was enveloped by a secular spin-down trend with the period reaching up to $530.430 \pm 0.014$ s at MJD 47221.97 \citep{Corbet1993}.



A four year long dedicated monitoring of 4U 1538-522 with Burst and Transient Source Experiment (BATSE) onboard \emph{Compton Gamma-Ray Observatory (CGRO)} revealed that the source underwent a torque reversal from spin-down phase to a secular spin-up phase sometime between 1988 and 1989 \citep{Rubin1997}. The source underwent another torque reversal to spin-down after about 20 yr as inferred from the \emph{INTEGRAL} and \emph{Fermi}/Gamma-ray Burst Monitor (GBM) spin-period observations \citep{Hemphill2013}.   

In addition to the temporal behavior, the spectral study of 4U 1538-522 is also interesting. The persistent broad-band energy spectrum of 4U 1538-522 is well described by an absorbed power-law modified by a high-energy cut-off, an Fe fluorescence emission line at 6.4 keV and a cyclotron resonance scattering feature (CRSF) around 22 keV. This implied a magnetic field strength of $1.7 (1+ {\rm z}) \times 10^{12}$ G \citep{Nagase1989,Clark1990}. The eclipse spectrum is consistent with a progressive covering of the primary Comptonization spectrum \citep{Clark1990,Robba2001,Rodes2009,Hemphill2013,Hemphill2014}. Several emission lines from highly ionized species of S, Si, and Mg below 3 keV have also been detected in the \emph{XMM Newton} spectrum of 4U 1538-522 \citep{Rodes2011}.



 Apart from these, a weak absorption feature has been detected in some of the spectral works. \citet{Robba2001} found it at around 51 keV (\emph{BeppoSAX}), \citet{Rodes2009} detected it at $\sim$ 47 keV (\emph{RXTE} and \emph{INTEGRAL}) and \citet{Hemphill2013} have reported this occurrence at $\sim$ 49 keV (\emph{Suzaku}). These energies are slightly higher than the first harmonic of CRSF. The CRSF energy shows a weak correlation with the source luminosity \citep{Hemphill2014}. A secular increase in the CRSF centroid energy has also been observed in 4U 1538-522 over 8.5 yr between \emph{RXTE} and \emph{Suzaku} observations \citep{Hemphill2016,Hemphill2019,Maitra2019}. 

 In this work, we present our results of the timing and broad-band spectral analysis of HMXB 4U 1538-522 with the \emph{NuSTAR} observations from 2019 and 2021 and \emph{Fermi}/GBM spin-period history.

\section{Observations}

\begin{figure*}
\centering
	\includegraphics[width=0.97\columnwidth]{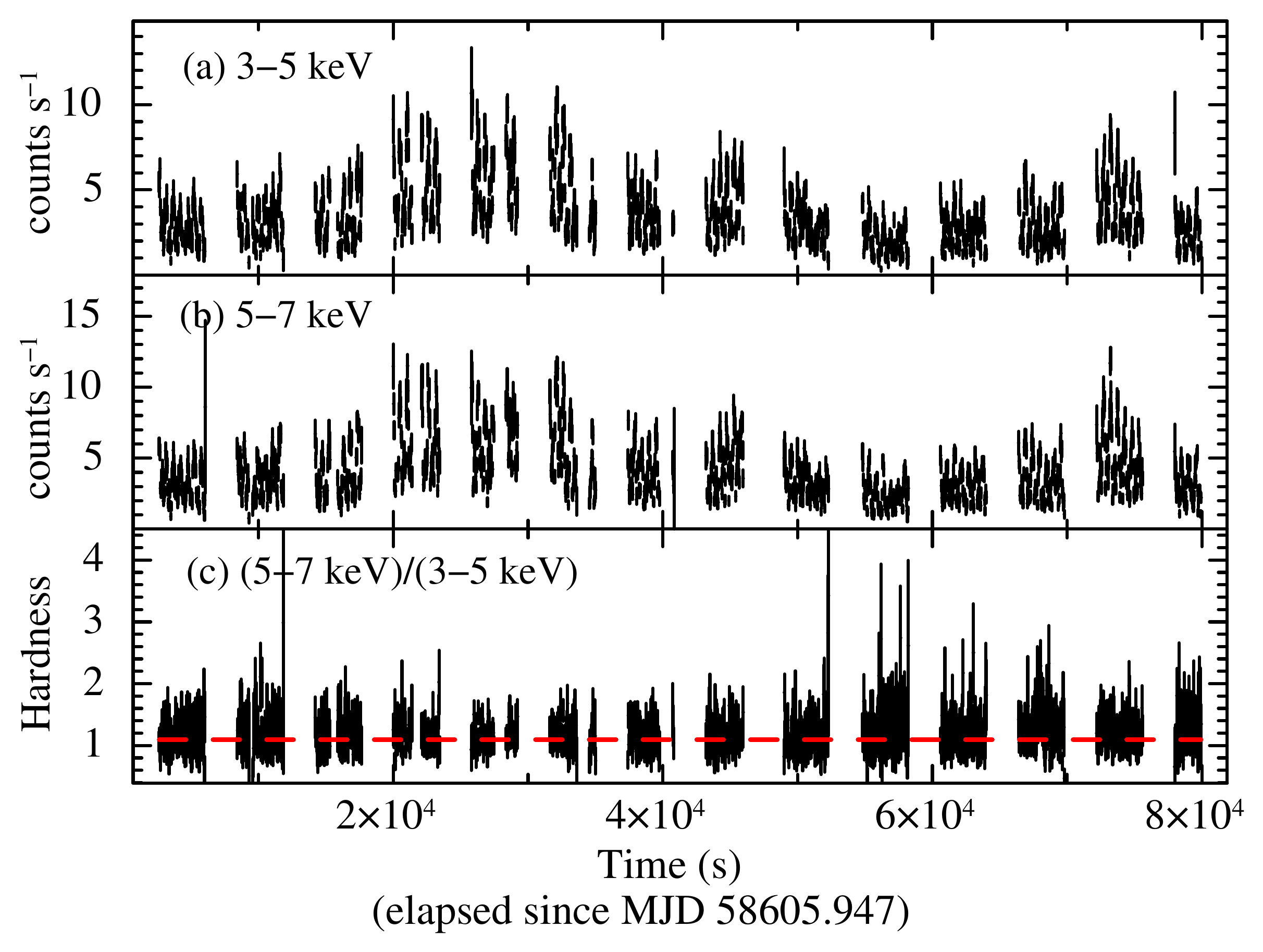}
 	\includegraphics[width=0.95\columnwidth]{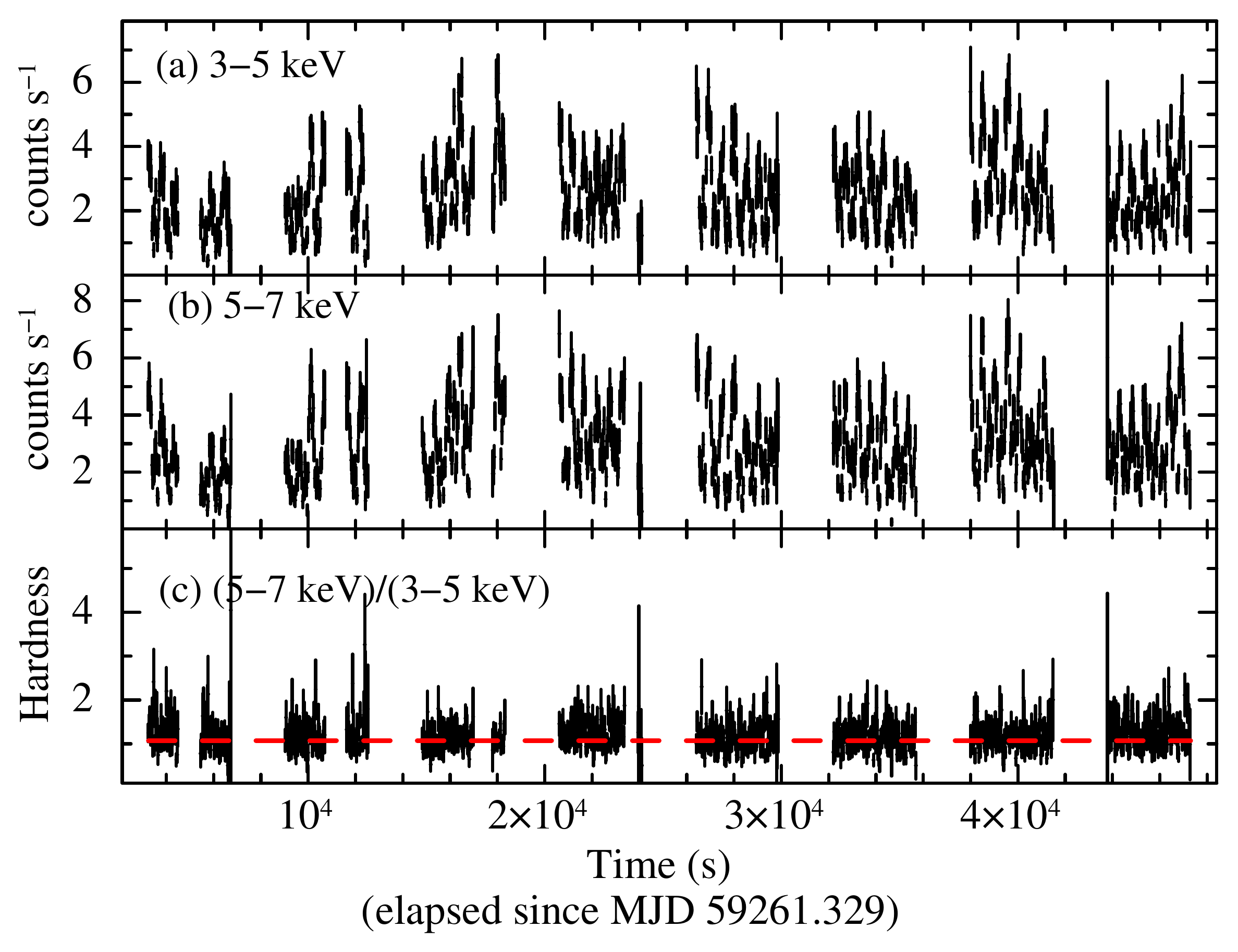}
	\includegraphics[width=0.97\columnwidth]{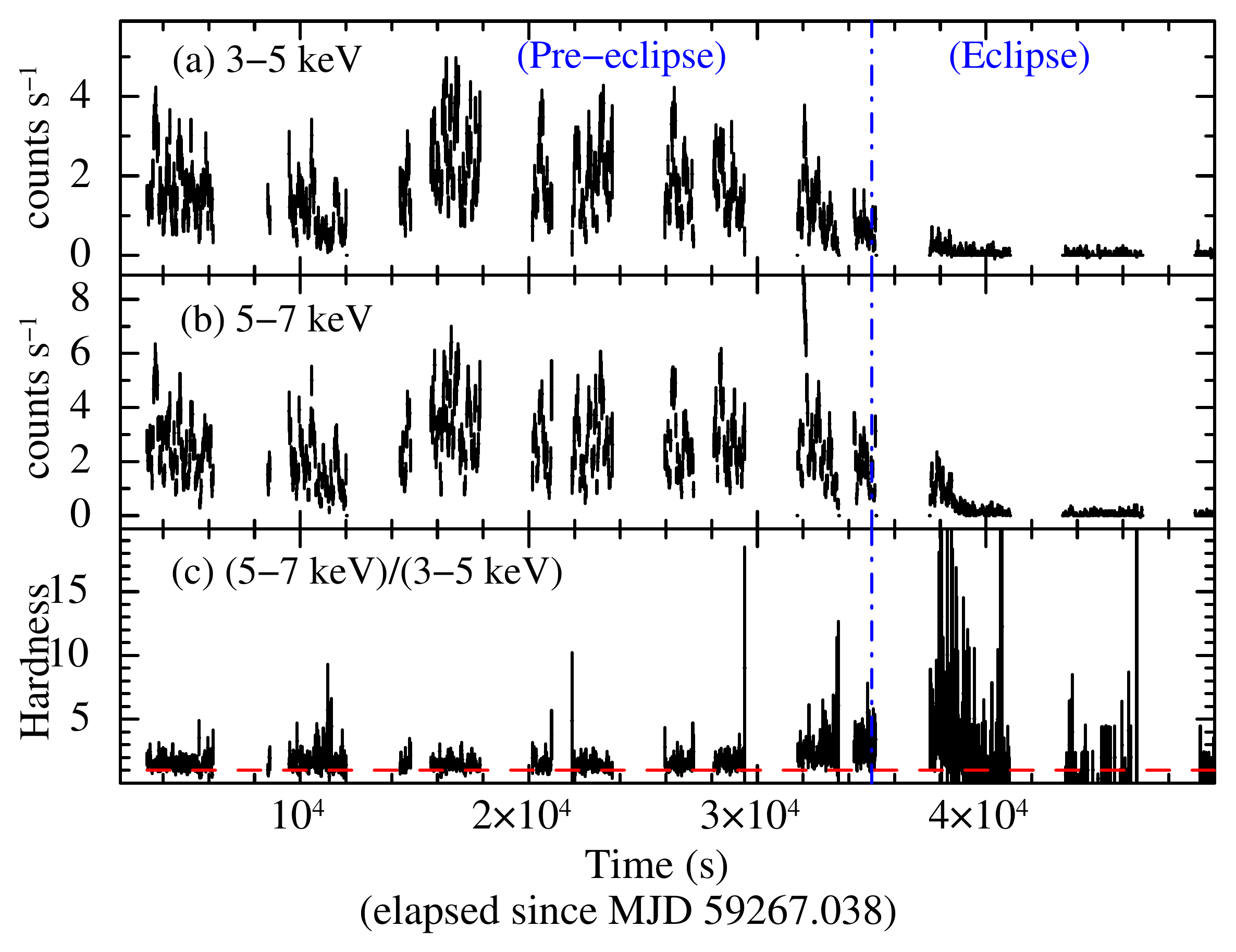}
    \caption{Background-subtracted \emph{NuSTAR}/FPMA light curves of 4U 1538-522 binned at 32 s for all three observations. Upper two panels in each plot show the energy-resolved light curves in 3--5 keV and 5--7 keV, respectively. The bottom panel gives the hardness ratio given as the ratio of 5--7 and 3--5 keV count rates. The red dashed horizontal line is the reference line at HR of 1. The blue dash-dot vertical line in the third plot divides the observation marks the beginning of the eclipse around the orbital phase of 0.93.} 
    \label{fig:light}
\end{figure*}

\emph{NuSTAR} is the first space mission by NASA, designed for the high-energy X-ray imaging purpose. Its payload is equipped with advanced X-ray focusing telescopes that provide good sensitivity at hard X-ray energies above 10 keV \citep{Harrison2013}. At the focal planes of the two co-aligned, grazing incidence telescopes are two identical detectors, focal plane module A (FPMA) and B (FPMB). The collective field of view of the detectors is 10 arcmin, with an angular resolution of 18 arcsec over the 3--70 keV instrumental band-pass. It provides an energy resolution of 400 eV at 10 keV and 900 eV at 60 keV.

4U 1538-522 was observed with \emph{NuSTAR} three times between 2019 and 2021 for a combined exposure of 80 ks per module. The log of observations used in this work is given in Table~\ref{tab:obslog}. We performed the standard data reduction of the raw data files by using the \textsc{nustardas}\_04May21\_v2.1.1 and the \emph{NuSTAR} \textsc{caldb} version 20211103, provided with the \textsc{heasoft} v6.29a. The level 2 event files were generated by applying the standard screening and calibration on the raw data by using the \textsc{nupipeline} task. For scientific product files, we used a circular regions of radii 80 arcsec centred on the source for both FPMA and FPMB, for all three observations. The background events were accumulated using a circular region with same radius away from the source, for each observations. We applied the solar system barycentric correction to the photon arrival times in all event files by using the source coordinates R.A. = 235.59734$^{\circ}$ and Dec. = -52.38599$^{\circ}$ \citep{Gaia2020}. We used the high-level task \textsc{nuproducts} to generate the averaged and energy-resolved light curves, source and background spectra and corresponding response files for each detector from all three observations. The barycenter-corrected light curves were further corrected for the orbital motion by using the latest orbital ephemeris of \citet{Hemphill2019}. 

The background-subtracted light curves for all three observations are shown in Figure~\ref{fig:light}, binned at 32 s. The top and middle panels in each plot correspond to the 3-5 and 5-7 keV energy ranges, respectively. The bottom panel corresponds to the hardness ratio (HR) for each observation, defined as the ratio of count rate in 5-7 keV and 3-5 keV. Based on the updated orbital ephemeris of \citet{Hemphill2019}, the three observations were taken during the orbital phases of 0.489--0.727, 0.272--0.409, and 0.803--0.948, respectively, with phase zero at the center of the eclipse. The last observation covers the ingress and a small part of the eclipse starting after 35 ks into the observation. 

Compared to the first observation, the X-ray intensity was almost 50 per cent less in the last observation. The intensity variation was marginal in individual observations with insignificant change in the HR across three observations, except for the last observation where HR was higher during the eclipse phase of the observation. The hardening of the emission spectra during eclipses has been observed in several other eclipsing sources (e.g. X 2127+119, \citet{Ioannou2002}; XTE J1710-281, \citet{Younes2009}; 2S 0921-63, \citet{Sharma2022c}).  For the spectral analysis, we have used the entire time-averaged data for the first two observations and resolved the data into two parts: pre-eclipse and eclipse (Fig.~\ref{fig:light}), for the last observation.

The source emission was detected up to 60 keV above background for the first two as well as the pre-eclipse segment of the third observation. The emission from source during the eclipse was dominant up to 50 keV only. Thus, we limit our spectral analysis in the energy range 3--60 keV for the first, second, and pre-eclipse spectra and in the range 3--50 keV for the eclipse spectrum. We re-binned the spectra from first and second observations to contain a minimum of 25 and 20 counts per energy bin, respectively, to improve the statistics. The pre-eclipse and eclipse spectra from the last observation were re-binned with 10 counts per energy bin.    

\section{Results}
\subsection{Timing analysis}

We used the orbital corrected, 3--60 keV light curves with a binning of 0.1 s from all the three observations to search for pulsations by using the epoch folding and $\chi^2$ maximization method \citep{Leahy1983} with the \texttt{ftool} \textsc{efsearch}. Pulsations were detected around 526 s in all three observations except during the eclipse phase of the third observation, where no pulsation was detected significantly. 

\begin{figure}

\includegraphics[width=\columnwidth]{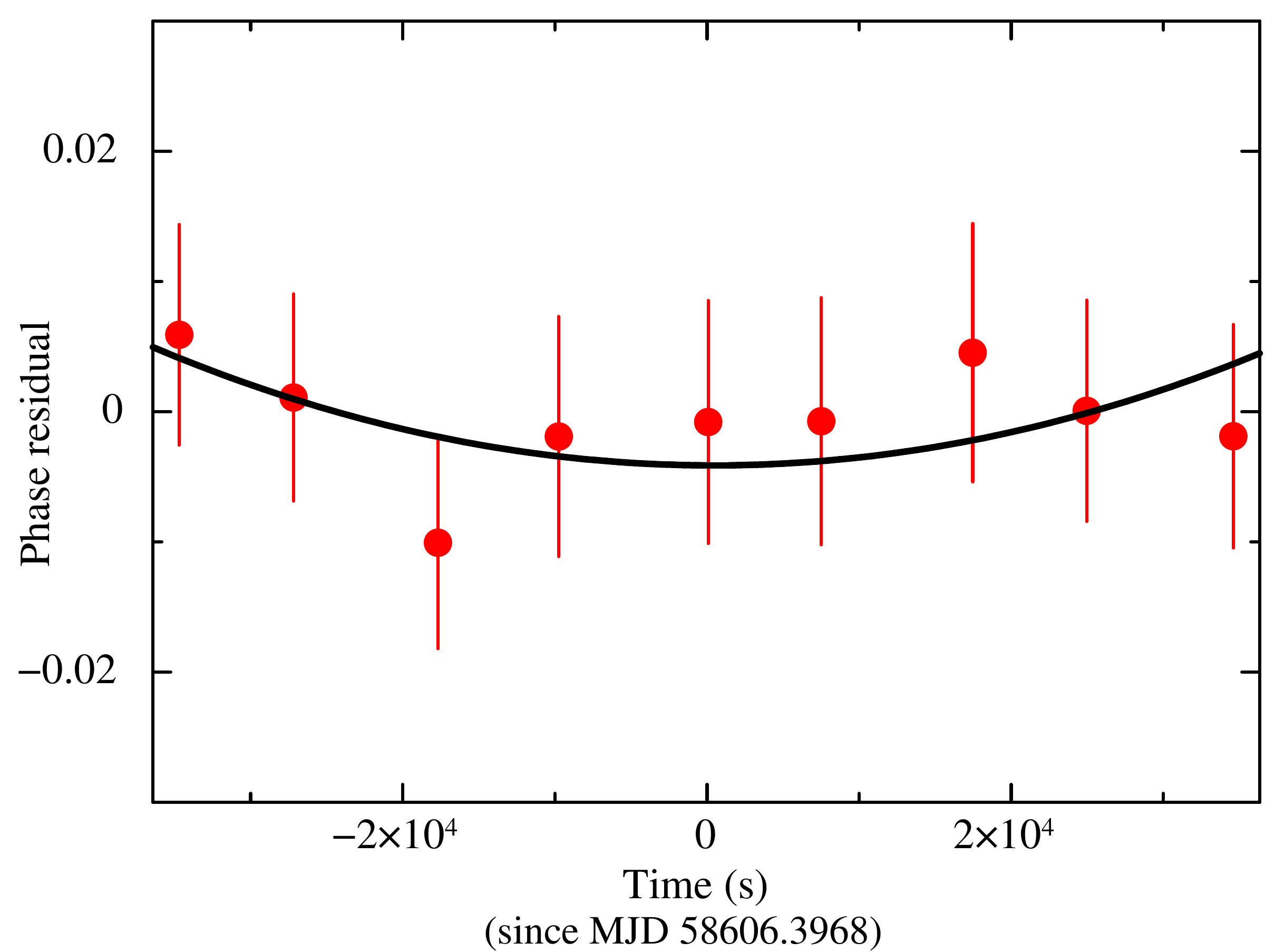}
\caption{Phase residual for the pulse-coherent timing analysis. The solid black curve corresponds to the best-fitting function up to a quadratic term.}
 \label{fig:toa}
\end{figure}

We performed the pulse-coherent timing analysis to estimate the rate of change of spin-period. Due to the short exposures and a large spin-period, the pulse-coherent analysis for the second and third observations failed to provide any results. For the first observation, we divided the entire light curve into 9 intervals of about 8200 s each (32 phase bins per period and 500 bins per interval) and cross-correlated the pulse profile from each interval with the averaged pulse profile over the entire light curve. The resulting phase delays and pulse arrival times were then fit to the equation~\ref{eq:phase} \citep{Deeter1981}, to obtain the best period and period derivative (Fig~\ref{fig:toa}). We obtained a pulse period of 526.57(6) s and a spin-down rate, $\dot{P} = 6.6_{-6.0}^{+2.4} \times 10^{-6}$ s s$^{-1}$ for the first observation. 
\begin{equation}
\phi(t) \ = \ \phi_{0} \ + \delta \nu (t-t_{0}) \ + \ \dot{\nu} (t-t_{0})^{2}/2
\label{eq:phase}
\end{equation}

\begin{table}
	\caption{Best-spin period values obtained for all three \emph{NuSTAR} observations of 4U 1538-522.}
	\label{tab:period}
	\begin{tabular*}{\columnwidth}{ccccc}
	\hline
        Obs\_ID & Epoch & Period  & $\dot{P}$ & Period$_{\rm GBM}^{a}$ \\
		      & (MJD) & (s) & (s s$^{-1}$) & (s) \\
		\hline
	   30401025002 & 58606.40 & 526.57(6) & $6.6_{-6.0}^{+2.4} \times 10^{-6}$ & 526.73(1) \\[0.5ex]
		 30602024002 & 59261.32 & 526.2(1) & - & 526.20(1)\\[0.5ex]
		 30602024004 & 59267.03 & 526.3(1) & - & 526.20(1) \\[0.7ex]
	\hline
    \multicolumn{5}{l}{\textit{Note.}$^{a}$ Period as measured with \emph{Fermi}/GBM closest to the observation epoch.}
 	\end{tabular*}
\end{table}

We used the method of \citet{Lutovinov2012} and \citet{Boldin2013} to evaluate the uncertainty in the estimated best-periods for the second and third observations by simulating 1000 light curves. The best-periods and the corresponding uncertainties are enlisted in Table~\ref{tab:period} along with the reference epochs. The estimated pulse period values for the three observations are in good agreement with the period measured by \emph{Fermi}/GBM \footnote{\url{https://gammaray.nsstc.nasa.gov/gbm/science/pulsars/lightcurves/4u1538.html}} \citep{Meegan2009} near the time of the observations.

\begin{figure*}

\includegraphics[width=\columnwidth]{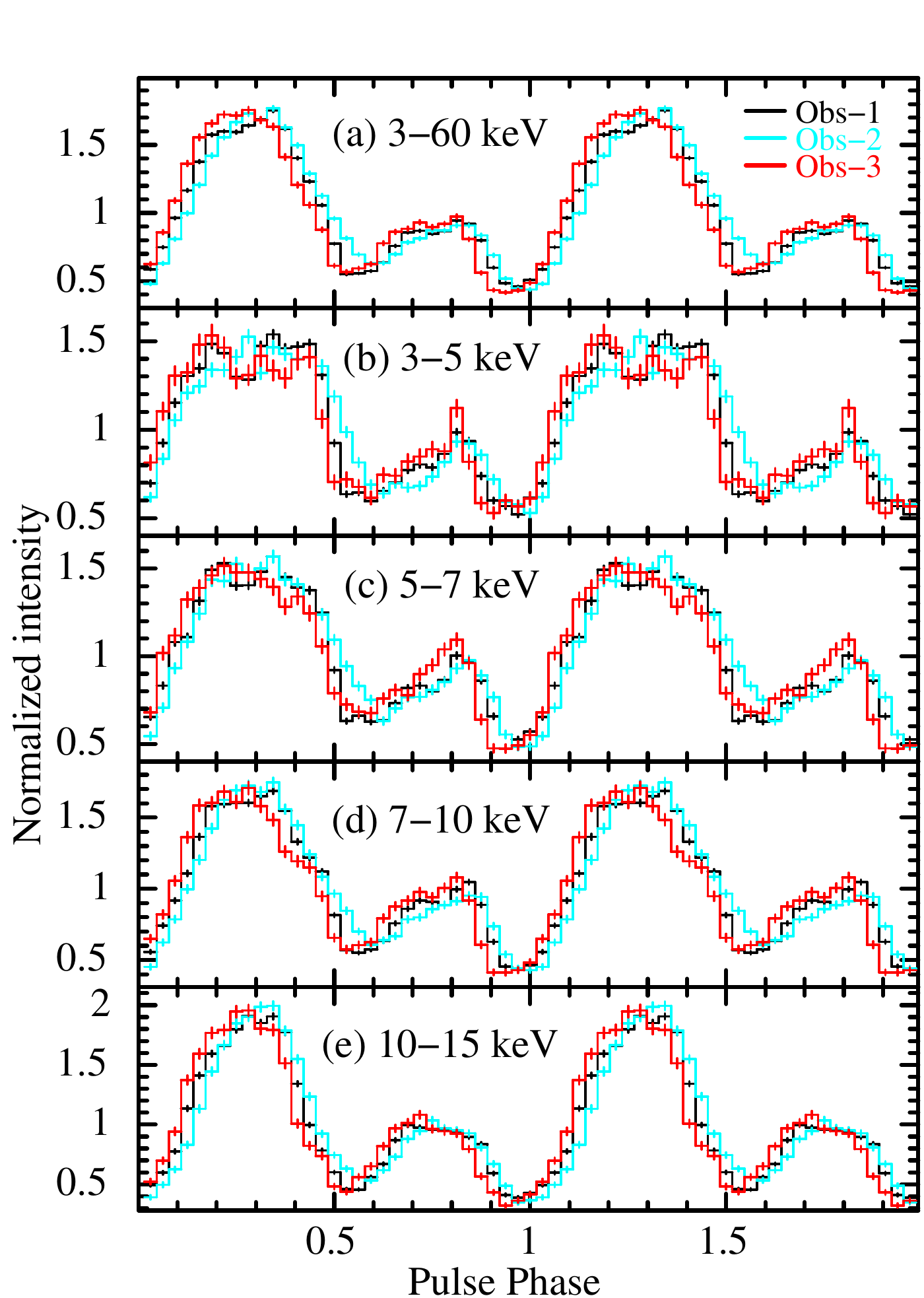}
\includegraphics[width=\columnwidth]{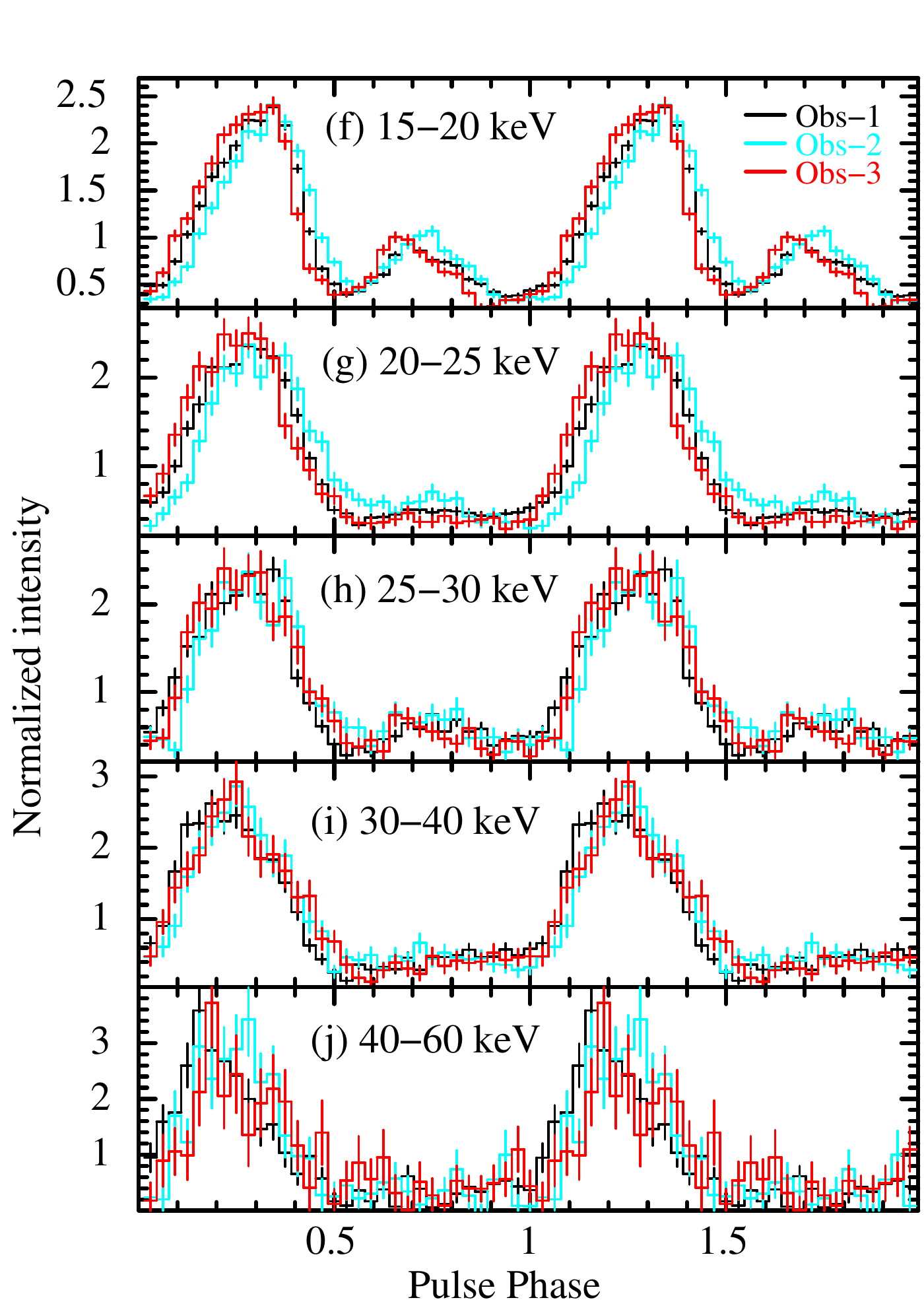}

\caption{Energy-resolved pulse profiles obtained from epoch-folding of \emph{NuSTAR} light curves with best-periods from Table~\ref{tab:period}, for observations 1 (black), 2 (blue), and 3 (red). The profiles are plotted for two cycles for clarity.}
 \label{fig:profile}
\end{figure*}
We also generated the energy resolved light curves in 3--5, 5--7, 7--10, 10--15, 15--20, 20--25, 25--30, 30--40, and 40--60 keV and folded the light curves at the respective best-periods for each observation to study the evolution of the profile with energy and time. All the pulse profiles for three observations are shown in Figure~\ref{fig:profile}. The pulse profile of 4U 1538-522 shows significant evolution with energy. The average 3--60 keV profile is double peaked with a broader strong peak and a weaker peak. For all the three observations, the primary peak is broader for energies below 7 keV with a small dip like structure near the maxima. The dip disappears and the peak becomes sharp above 10 keV and shifts slightly in the phase above 25 keV with a decrease in its width. The second weaker peak has a flat top in the average profile. Up to 10 keV, it is sharp and it has a marginal detection in 25--30 keV energy range. It disappears thereafter. 


\subsection{Long-term period evolution and torque reversals}
Wind-fed pulsars often show random, short-term fluctuations in their spin-period \citep{Deeter1989,Bildsten1997} with occasional long-term torque reversals (e.g. Vela X1, \citet{Hayakawa1982}; GX 1+4, \citet{Gonzalez2012}; OAO 1657-415, \citet{Sharma2022a}). 4U 1538-522 is one of such sources which show similar short-term variations and torque reversals \citep{Makishima1987,Rubin1997}. Due to lack of continuous monitoring data, the two reported torque reversals of 4U 1538-522 have only provided tentative epochs for the reversals. 

\begin{figure}
\centering
	\includegraphics[width=\columnwidth]{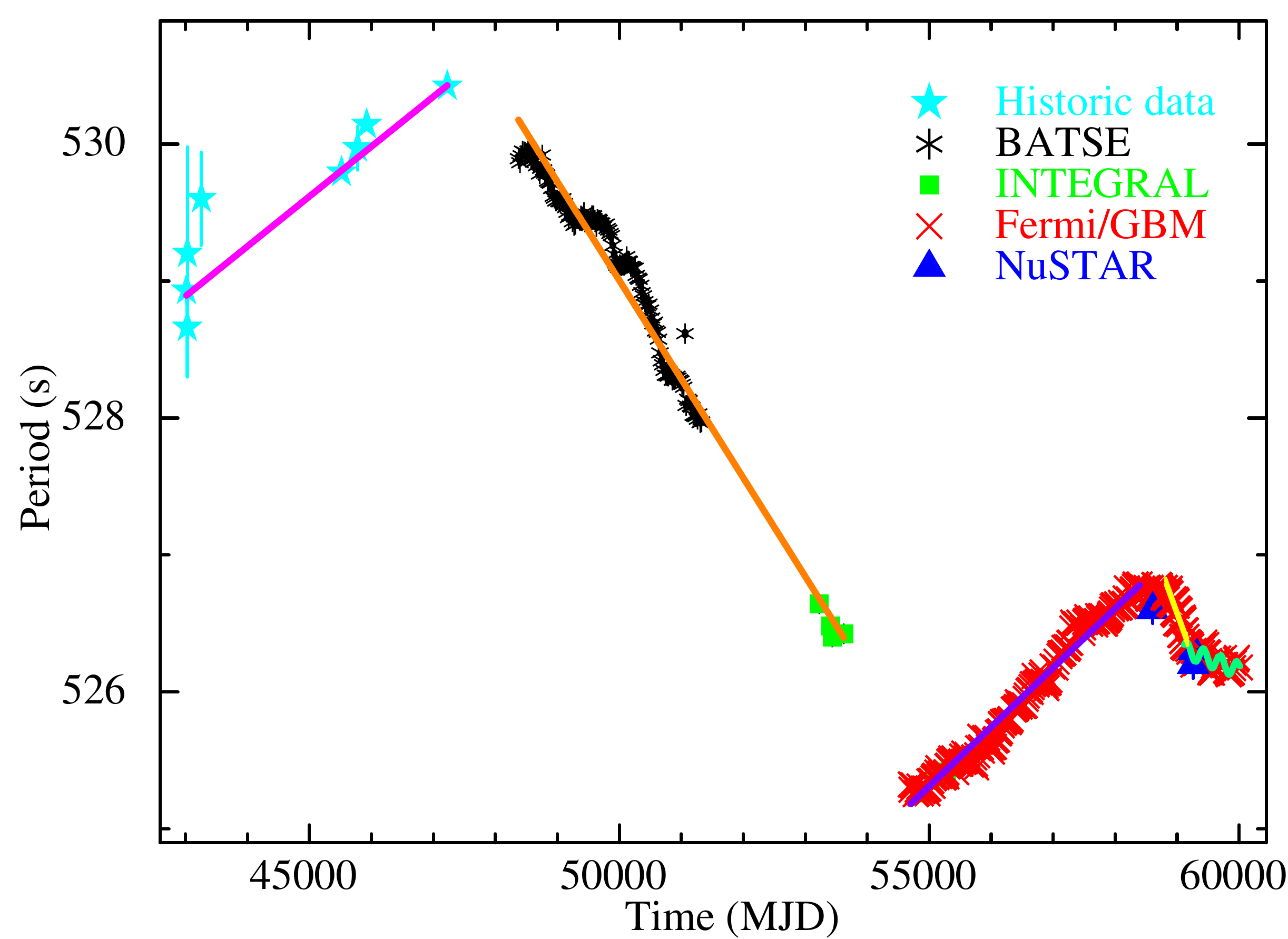}
\caption{Long-term spin period history of 4U 1538-522, spanning between 1976 and 2022. The different data sets are represented by different markers. Historic data from earlier missions are marked with light blue star. Black asterisk corresponds to \emph{CGRO}/BATSE monitoring data \citep{Rubin1997}. Filled green square represents the \emph{INTEGRAL} results from \citet{Hemphill2013}. Data marked in red crosses correspond to the \emph{Fermi}/GBM archival data and filled blue triangle represent the results from this work. Solid lines corresponds to the best linear fit for each data group.} 
 \label{fig:period}
 \end{figure}

Spin-period history of 4U 1538-522 is available from as early as August 1976 from various missions. \citet{Hemphill2013} presented the spin-period evolution of the source covering a time-line of about 36 yrs between 1976 and 2012 and suggested a torque-reversal tentatively around 2009. We have extended the time-line by another 10 yr, by including the regular monitoring data from \emph{Fermi}/GBM. The complete spin-period history of 4U 1538-522 is shown in Figure~\ref{fig:period} between August 1976 and November 2022, including results from the present work. 


\begin{table*}
\caption{Log of the spin-period history of 4U 1538-522 with different missions.}
 	\label{tab:pderiv}
\begin{tabular}{cc|cc|ccc}
\hline
S.No. & Group & \multicolumn{2}{c}{Time line} & Duration & $\dot{P}^{a}$ & Reversal$^{b}$ \\
& & (MJD) & (dd:mm:yyyy) & (yr) & ($10^{-9}$ s s$^{-1}$) & (MJD)\\
 \hline
1 & Historic data & 43016.3--47223.0 & 26:08:1996--03:03:1988 & 11.5 & 4.2(2) & 47750 \\ [0.5ex]
\hline
2 & \emph{CGRO}/BATSE & 48372.9--51325.2 & 26:04:1991--27:05:1999 & 8 & -8.3(1) & \\[0.5ex]
& & & & & \\
   &  \emph{INTEGRAL} & 53222.1--53620.8 & 05:08:2004--07:09:2005 & 1 & & \\[0.5ex]
 \hline

3 & \emph{INTEGRAL} & 54863.1--55276.8 & 24:01:2009--02:04:2010 & 1 & 4.9(1) & 54882 - 55080 \\[0.5ex]
& & & & &\\
   & \emph{Fermi}/GBM & 54703.4--58796.7 & 25:08:2008--09:11:2019 & 11 &  \\[0.5ex] \hline

4 & \emph{Fermi}/GBM & 58811.5--59102.5 & 24:11:2019--10:09:2020 & 1 & -14.4(3) & 58800\\
& & & & &\\
   &  & 59115.4--60012.8 & 23:09:2020--09:03:2023 & 2.5 & -1.9(1)\\
\hline
\multicolumn{7}{l}{\textit{Notes.}$^{a}$ Period derivative evaluated from the linear fitting of the data.}\\
\multicolumn{7}{l}{$^{b}$ Based on the extrapolation of the linear fittings.}

\end{tabular}
\end{table*}

Including all the available period history and recent \emph{Fermi}/GBM data, the period derivatives have been estimated for different spin phases by fitting the data with a linear function. The best-fitting results are plotted in the Figure~\ref{fig:period} with solid lines and the \textbf{estimated} parameters are reported in the Table~\ref{tab:pderiv}. 

 \begin{figure}
\centering
	\includegraphics[width=\columnwidth]{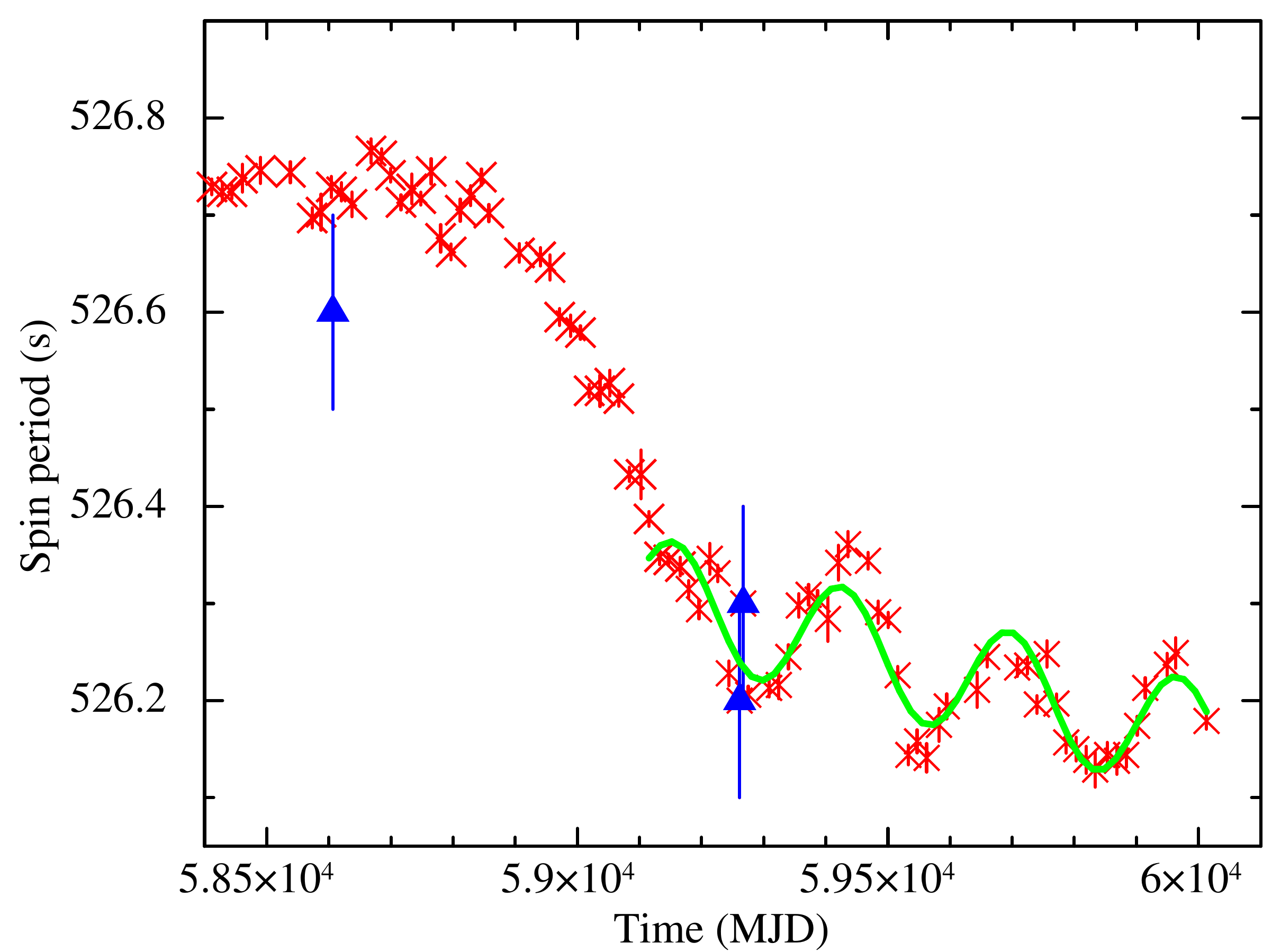}
\caption{Recent spin-period trends of 4U 1538-522 with the \emph{Fermi}/GBM data. Solid green curve represents the best-fitting linear and sinusoidal model for data starting from MJD 59115.} 
 \label{fig:sine}
 \end{figure}

The entire spin-period history can be divided into five groups. The first group comprises of the historic data covering 12 yr between 1976 and 1988 (MJD 43016.3--47223) with spin-estimates from \emph{OSO-8} \citep{Becker1977}, \emph{Ariel V} \citep{Davison1977a}, \emph{Tenma} \citep{Makishima1987}, \emph{Exosat} \citep{Cusumano1989}, and \emph{Ginga} \citep{Nagase1989} (light blue star in Fig~\ref{fig:period}). Based on the linear fitting results, we estimated a spin-down rate of $\dot{P} = 4.2(2) \times 10^{-9}$ s s$^{-1}$ for this period. This value is consistent with reported result of $3.9 \times 10^{-9}$ s s$^{-1}$ by \citet{Cusumano1989} and \citep{Robba1992} for the same duration.

The second group includes the \emph{CGRO}/BATSE monitoring data of the source from April 1991 (MJD 48372.8) to May 1999 (MJD 51325), covering a time span of about 8 yr (black asterisk) and \emph{INTEGRAL} data \citep{Hemphill2013} from August 2004 to May 2005 (MJD 53222.1--53620.8) (filled green square). The source exhibited a long-term spin-up trend with random, short-fluctuation superimposed on it. Our best-fitting yields a spin-up rate of $-8.3(1) \times 10^{-9}$ s s$^{-1}$ for this duration. The extrapolation of the linear trends for this data group and the data for previous spin-down phase gives the epoch for the first torque reversal somewhere around August 1989 ($\sim$ MJD 47750). \citet{Rubin1997} suggested the epoch for this torque reversal sometime in 1988 and a spin-period derivative of about $-5 \times 10^{-9}$ s s$^{-1}$ for their \emph{CGRO} data.

The third group includes the remaining \emph{INTEGRAL} data \citep{Hemphill2013} from January 2009 to April 2010 (MJD 54863.1--55276.8) along with the first 11 yr archival data from \emph{Fermi}/GBM (red crosses) between August 2008 (MJD 54703.4) and November 2019 (MJD 58796.7). The \emph{INTEGRAL} data points are not visible as they coincides with the \emph{Fermi} data. For this duration, the period trend shows that the source was spinning down since the beginning of the monitoring. This spin-down phase continued for about 10 yr till late 2018 ($\sim$ MJD 58400) and after that it was effectively constant for almost an year till late 2019 ($\sim$ MJD 58800). The linear fit to the spin-down phase data till late 2018 gives a period derivative, $\dot{P} = 4.9(1) \times 10^{-9}$ s s$^{-1}$. The extrapolation of the previous and this data group trend suggests the epoch of torque reversal from the earlier spin-up phase, sometimes between February and September 2009 ($\sim$ MJD 54882--55080). \citet{Hemphill2013} also suggested the epoch for this second torque reversal to be sometime in 2009. However, they reported a relatively higher spin-period derivative of $\sim 9.5 \times 10^{-9}$ s s$^{-1}$ for the first four year data up to late 2012.

Around MJD 58800, the source underwent a torque reversal for the third time and started spinning up once again. The last data group corresponds to the \emph{Fermi}/GBM data, covering 3.5 yr between November 2019 and March 2023 (MJD 58811.5--60012.8). The overall trend shows that the source is spinning up. However, the current spin period trend shows a sinusoidal variation after September 2020 ($\sim$ MJD 59110). We obtained a spin-up rate of $\dot{P} = -14.4(3) \times 10^{-9}$ s s$^{-1}$ for the period between November 2019 (MJD 58811.5) and September 2020 (MJD 59102.5). In order to estimate the period for the latest sinusoidal variation, we fit the data between MJD 59115.4 and MJD 60012.8 with a linear function modulated by a sinusoidal function. The best-fitting yields a period of $271.2 \pm 2.6$ d and an amplitude of variation of $0.059 \pm 0.004$ s with an underlying spin-up rate of $-1.9(1) \times 10^{-9}$ s s$^{-1}$. The solid green curve in Figure~\ref{fig:sine} shows the best-fitting linear and a sinusoidal model.

\subsection{Spectral analysis}
We have performed the simultaneous spectral fitting of the \emph{NuSTAR} FPMA and FPMB spectra of 4U 1538-522 by using the \textsc{xspec} v12.12.0 \citep{Arnaud1996} for each observation individually. We introduced a cross-normalization constant ($C_{\rm FPMB}$) for the simultaneous fitting to account for the different detector responses. We froze its value at 1 for FPMA and left it as free parameter for FPMB. We obtained a value of $0.98 \pm 0.01$ for the first observation and a value of $1.02 \pm 0.01$ for the second and third observations. We have used the updated solar abundances from \citet{Wilms2000} and the photoelectric cross-sections of \citet{Verner1996} for the spectral fittings.

Several empirical models can be used to model the X-ray continua of accreting HMXBs. We used three different forms of power law model, i.e., modified by a high-energy cut-off (HEPL), a cut-off power law (CPL), and broken power law (BKPL). We added a Gaussian (\texttt{gaus}) for the Fe K$_{\alpha}$ feature and included a multiplicative absorbed Gaussian component (\texttt{gabs}, in \textsc{xspec}) to account for the CRSF feature around 22 keV to the models. All the models failed to provide an acceptable fitting with significant residual near lower energy along with a high energy tail. The addition of \texttt{tbabs} component to account for the photoelectric absorption due to the interstellar medium, did not provide any significant improvement. While the column density parameter was poorly constrained for HEPL, it was not favoured by either CPL or BKPL. The attempt to fix the value of column density at values from literature introduced systematic residual at the lower energy end and it worsened the fitting. We have therefore, not included this component in our subsequent analysis. We added a blackbody component, \texttt{bbodyrad} to account for the low-energy residual. This improved the fitting statistically with $\chi^2 /\nu$ of 1668/1521, 1757/1523, and 1726/1522 with the best-fitting CRSF energy of $21.83 \pm 0.11$, $22.24 \pm 0.11$, and $21.07 \pm 0.16$ for HEPL, CPL, and BKPL models, respectively for the first observation. However, the residuals near high-energy persisted along with a hint of weak, dip-like feature around 30 keV for all three models.


\begin{figure}
\centering
	\includegraphics[height=\columnwidth,width=\columnwidth]{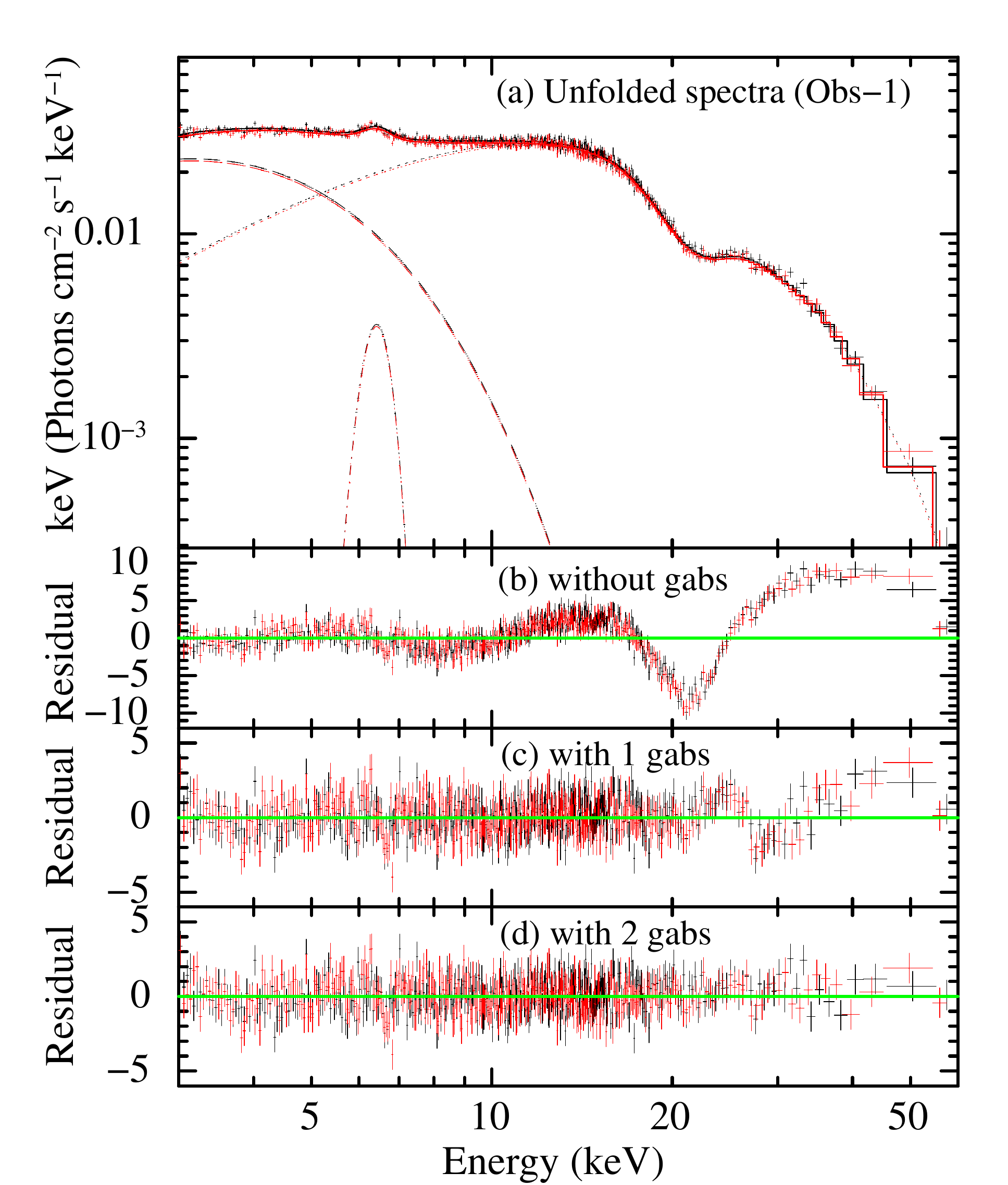}
\caption{(a) Best-fitting unfolded \emph{NuSTAR} FPMA (black) and FPMB (red) spectra of 4U 1538-522 for the first observation with the CPL model M1. (b) Residual without any \texttt{gabs} component. (c) Residual after including a \texttt{gabs} component at 22 keV. (d) Residual after including a second \texttt{gabs} component. The dashed, dotted, and dash-dotted lines represent \texttt{bbodyrad}, cut-off powerlaw, and \texttt{gaus} components, respectively.} 
 \label{fig:cpl1}
 \end{figure}


\begin{figure}
\centering
	\includegraphics[height=\columnwidth,width=0.95\columnwidth]{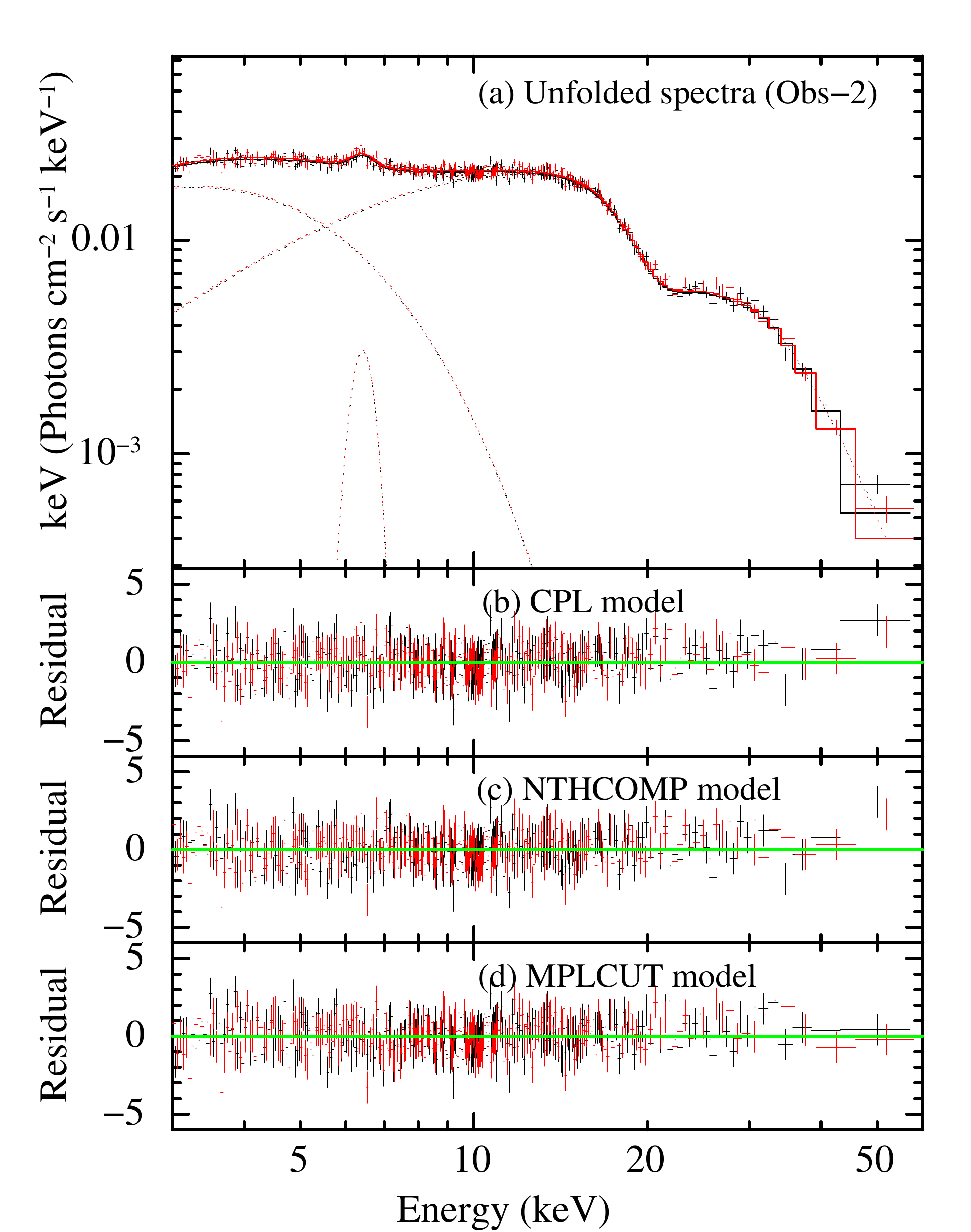}
\caption{Best-fitting FPMA (black) and FPMB (red) spectra of 4U 1538-522 for the second observation. (a) Unfolded spectra and model components. (b) Residual with CPL model M1 (c) Residual with \texttt{nthcomp} model M2. (d) Residual with \texttt{mplcut}  model M3. The dashed, dotted, and dash-dotted lines represent \texttt{bbodyrad}, \texttt{cutoffpl}/\texttt{nthcomp}/\texttt{mplcut}, and \texttt{gaus} components, respectively.} 
 \label{fig:spec2}
 \end{figure}

While multiple CRSFs are commonly found in the spectra of HMXBs, these features appear at harmonically spaced energies. We included a second \texttt{gabs} component at 30 keV to model the dip-like feature in the spectra. The addition of second absorbed Gaussian component improved the fitting significantly for CPL ($\Delta \chi^2 = 120$) and BKPL ($\Delta \chi^2 = 46$) for a change of 3 degrees of freedom (dof). However, the improvement was marginal for HEPL ($\Delta \chi^2 = 11$ for $\Delta$ dof = 3) along with some residual at the higher energy end. While all the three models provided statistically similar fitting, CPL model provided relatively better fitting in terms of excess residual near the high-energy end. Therefore, we only retain the final CPL model, \texttt{gabs*gabs*(bbodyrad + cutoffpl + gaus)} (M1, hereafter) for our subsequent analysis. Figure~\ref{fig:cpl1} shows the best-fitting spectra from observation 1 with model M1, before and after adding the \texttt{gabs} components. The best-fitting parameter values are reported in Table~\ref{tab:spec1}. All the parameter uncertainties are at 90 per cent confidence level.


\begin{table*}
\caption{Best-fitting spectral parameters for the time-averaged \emph{NuSTAR} spectrum of 4U 1538-522 for the first observation. All the errors are quoted at 90 per cent confidence level.}
\label{tab:spec1}
\begin{tabular*}{1.9\columnwidth}{c c | c c | c c | c c } 

\hline
& & & & & & & \\
\textbf{Component} & \textbf{Parameters} & \multicolumn{2}{c}{\textbf{Model M1}} & \multicolumn{2}{c}{\textbf{Model M2}} & \multicolumn{2}{c}{\textbf{Model M3}} \\
\hline
 & & & & & & & \\
TBABS & $N_{\rm H}$ ($10^{22}$ cm$^{-2}$) &\multicolumn{2}{c}{$ - $} &\multicolumn{2}{c}{$ - $} & \multicolumn{2}{c}{$1.1 \pm 0.3$}   \\[0.7ex]

EDGE & $E_{\rm Fe \textsc{i}}$ (keV) & $-$ & $-$& $-$&$-$ & $ 7.25 \pm 0.08$ & $7.25 \pm 0.08$ \\[0.5ex]
 & $ \tau$ &$-$ & $-$& $-$& $-$& $ 0.07\pm 0.01$ & $0.07\pm 0.01$ \\[0.7ex]

BBODYRAD & $kT_{\rm BB}$ (keV) & $ 1.10 \pm 0.02 $ & $ 1.11 \pm 0.03 $ & $1.12_{-0.04}^{+0.02} $  & $ 1.19 \pm 0.02 $ & $-$ & $-$ \\[0.5ex]
 & $N_{\rm BB}$ & $ 11.97_{-0.62}^{+0.68}$ & $ 11.46_{-0.61}^{+0.64}$ & $9.62_{-0.30}^{+0.48}$ & $ 9.10_{-0.45}^{+0.58}$ & $-$ & $ - $ \\[0.7ex]
& ${R_{\rm BB}^a}$ (km) & $2.2 \pm 0.1$  & $2.2 \pm 0.1$  & $1.9 \pm 0.1$ & $1.9 \pm 0.1$ & $-$ & $-$ \\[0.5ex]
& $f^{a}_{\rm bol}$ & \multicolumn{2}{c}{$1.90 \pm 0.01$} & \multicolumn{2}{c}{$1.41 \pm 0.01$} & \multicolumn{2}{c}{$-$} \\[0.7ex]
 
 CPL & $\Gamma$ & $ -1.23_{-0.11}^{+0.10}$ & $ -1.12_{-0.19}^{+0.10}$ & $1.18_{-0.02}^{+0.04} $ & $ 1.22_{-0.07}^{+0.03}$ & $1.15 \pm 0.01$ & $1.15 \pm 0.01$ \\ [0.5ex]
  /NTHCOMP & $E_{\rm cut}$ (keV) & $ 5.32 \pm 0.14 $ & $ 5.61_{-0.17}^{+0.15} $ &-&-& $16.51_{-0.65}^{+0.54}$& $ 16.92_{-0.36}^{+0.49}$ \\[0.5ex]
  /MPLCUT & $kT_{\rm e}$ (keV) &$-$ & $-$ & $ 4.68_{-0.05}^{+0.06} $ & $ 4.94_{-0.13}^{+0.07} $ & $-$ & $-$ \\[0.5ex]
  & $E_{\rm fold}$ (keV) &$-$ &$-$ & $-$ & $-$ &$ 10.0 \pm 0.3 $  & $9.9_{-0.3}^{+0.2} $ \\[0.5ex]
  & ${\tau}$ &$-$ &$-$ & $22.5_{-2.5}^{+1.6}$ & $19.5_{-1.5}^{+5.0}$ & $-$ & $-$ \\[0.5ex]
 & Norm ($10^{-3}$) & $1.1 \pm 0.2$ & $1.2_{-0.4}^{+0.3}$ & $2.5_{-0.4}^{+0.7}$ & $3.3_{-1.1}^{+0.5}$ & $41.5 \pm 1.3 $ & $41.7 \pm 1.3$ \\[0.5ex]
 & $f^{a}_{\rm bol}$ & \multicolumn{2}{c}{$9.50 \pm 0.03$} & \multicolumn{2}{c}{$10.26 \pm 0.03$} & \multicolumn{2}{c}{$ 12.68 \pm 0.02$} \\[0.7ex]

 GAUS & $E_{\rm line}$ (keV) & $ < 6.41 $ & $< 6.41$ & $< 6.41 $ & $ < 6.41 $ & $< 6.41$ & $ < 6.41 $\\[0.5ex]
 & $\sigma$ (keV) & $0.37 \pm 0.07 $  & $0.34 \pm 0.07 $ & $ 0.33_{-0.06}^{+0.08} $ & $0.33 \pm 0.08 $ & $0.01^{\dag} $ & $0.01^{\dag}$ \\[0.5ex]
 & Norm ($10^{-4}$) & $5.2_{-0.8}^{+0.9}$ & $4.8_{-0.7}^{0.8}$ & $4.7_{-0.6}^{+0.9}$ & $24.6_{-0.7}^{+0.8}$ & $1.6 \pm 0.3 $ & $ 1.6 \pm 0.3 $ \\[0.5ex]
 & EW (eV) & $111_{-13}^{+19}$ & $102_{-13}^{+15} $ & $100_{-12}^{+16}$ & $98_{-12}^{+16}$ & $33_{-6}^{+5} $ & $33_{-5}^{+6}$\\[0.7ex]
 & $f^{b}_{\rm bol}$ & \multicolumn{2}{c}{$0.49 \pm 0.04$} & \multicolumn{2}{c}{$0.47 \pm 0.04 $} & \multicolumn{2}{c}{$ 0.17\pm0.02 $} \\[0.7ex]

 GABS & $E_{\rm line}$ (keV) & $ 22.24 \pm 0.11$ & $ 21.41_{-0.26}^{+0.45}$ & $22.33 \pm 0.10 $ & $ 21.42_{-0.21}^{+0.29}$ & $ 21.7 \pm 0.2 $ & $ 21.8_{-0.8}^{+0.3}$ \\[0.5ex]
 & Width (keV) & $3.38 \pm 0.13$ & $2.92_{-0.28}^{+0.24}$ & $3.56 \pm 0.10 $ & $2.96 \pm 0.23$ &  $ 3.15_{-0.21}^{+0.26} $ & $2.62_{-0.58}^{+0.28}$\\[0.5ex]
 & Strength & $ 6.14_{-0.34}^{+0.37}$ & $ 4.64_{-1.36}^{+1.27}$  & $ 6.85_{-0.27}^{+0.28}$ & $ 4.60_{-1.06}^{+1.37}$ & $ 5.04_{-0.38}^{+0.68}$ & $ 4.45_{-2.04}^{+0.41}$ \\[0.7ex]

 GABS & $E_{\rm line}$ (keV) & $-$ & $27.53_{-2.76}^{+2.01}$ & $-$ & $27.62_{-1.52}^{+1.92}$ & $-$ & $27.63_{-3.35}^{+0.60}$ \\[0.5ex]
 & Width (keV) & $-$ & $4.51_{-1.86}^{+1.70}$ & $-$ & $ 5.45_{-1.86}^{+1.39}$ & $-$ & $1.5_{-0.7}^{+2.0} $\\[0.5ex]
 & Strength & $-$ & $ 2.97_{-1.51}^{+2.42}$ & $-$ & $4.44_{-1.98}^{+1.71}$ & $-$ & $0.55_{-0.27}^{+2.04}$ \\[0.5ex]

 & & & & & & &\\
 &$f^{a}_{\rm Total}$ & \multicolumn{2}{c}{$11.44 \pm 0.02$} & \multicolumn{2}{c}{$11.72 \pm 0.30 $} & \multicolumn{2}{c}{$ 12.70 \pm 0.02  $} \\[0.7ex] 

\hline
 & $\chi^2$/dof & 1746.2/1523 & 1637.1/1520 & 1777.3/1523 & 1637.1/1520 & 1642.2/1520 &  1617.1/1517 \\
\hline

\multicolumn{8}{l}{\textit{Notes.} Model M1 = \texttt{gabs*gabs*(bbodyrad + cutoffpl + gaus)}} \\
\multicolumn{8}{l}{Model M2 = \texttt{gabs*gabs*(bbodyrad + nthComp[bb] + gaus)}} \\
\multicolumn{8}{l}{Model M3 = \texttt{tbabs*edge*gabs*gabs*(powerlaw*highecut + gaus + gaus)}}\\ 
\multicolumn{8}{l}{$^{a} f$ is the unabsorbed bolometric flux in 0.1--100.0 keV in $10^{-10}$ erg cm$^{-2}$ s$^{-1}$.}\\ 
\multicolumn{8}{l}{$^{b} f$ is the unabsorbed bolometric flux in 0.1--100.0 keV in $10^{-11}$ erg cm$^{-2}$ s$^{-1}$.}\\
\multicolumn{8}{l}{$^{\dag}$ Parameters fixed at the value for fitting.}
\end{tabular*} 
\end{table*}

\begin{figure}
\centering
	\includegraphics[height=\columnwidth,width=\columnwidth]{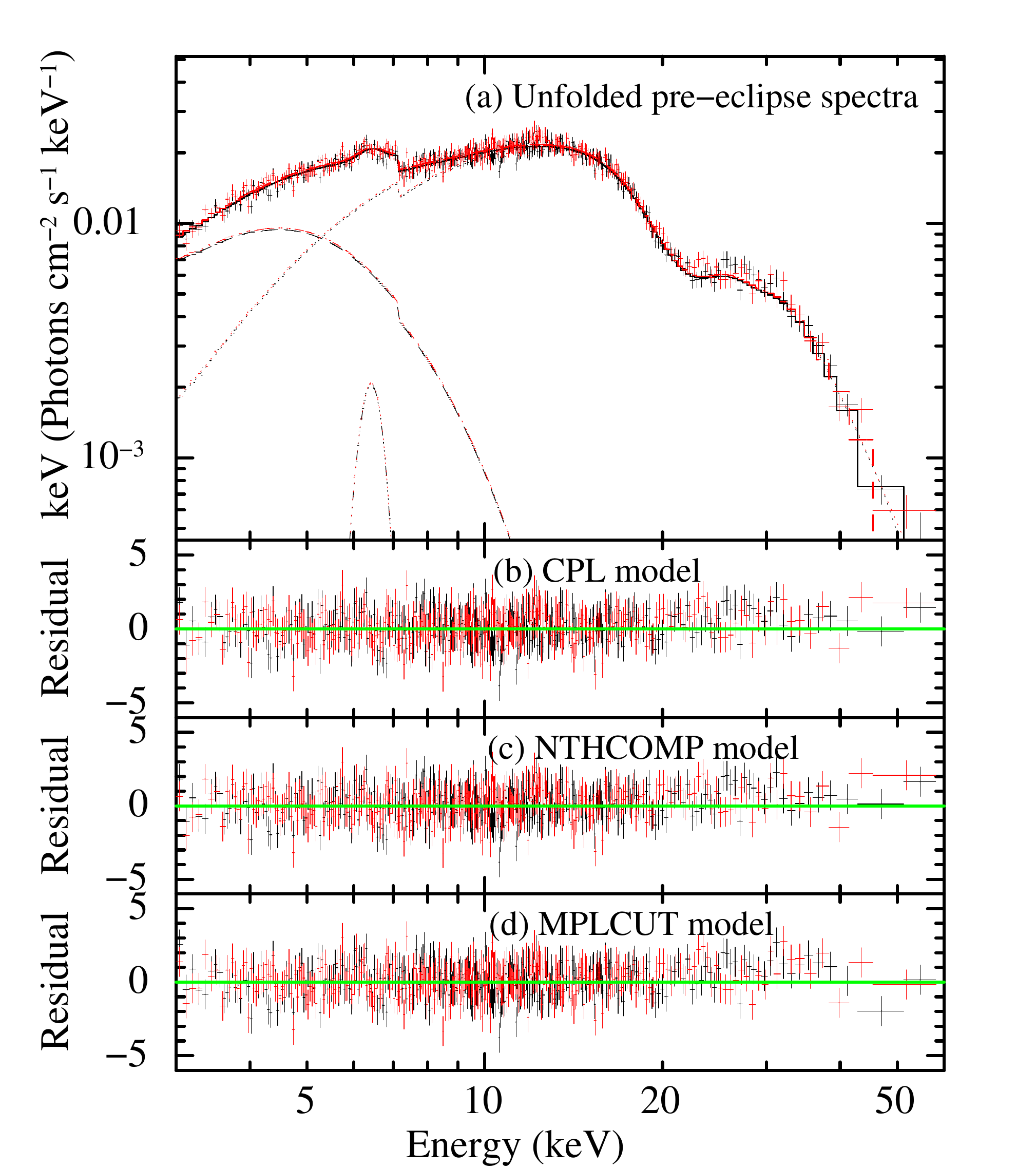}
\caption{Best-fitting FPMA (black) and FPMB (red) pre-eclipse spectra for the last observation. (a) Unfolded spectra and model components. Residual for models M1 (b), M2 (c) modified with \texttt{pcfabs} and M3 (d). The dashed, dotted, and dash-dotted lines represent \texttt{bbodyrad}, \texttt{cutoffpl}/\texttt{nthcomp}/\texttt{mplcut}, and \texttt{gaus} components, respectively.}
 \label{fig:spec3a}
 \end{figure}

\begin{figure}
\centering
	\includegraphics[height=\columnwidth,width=\columnwidth]{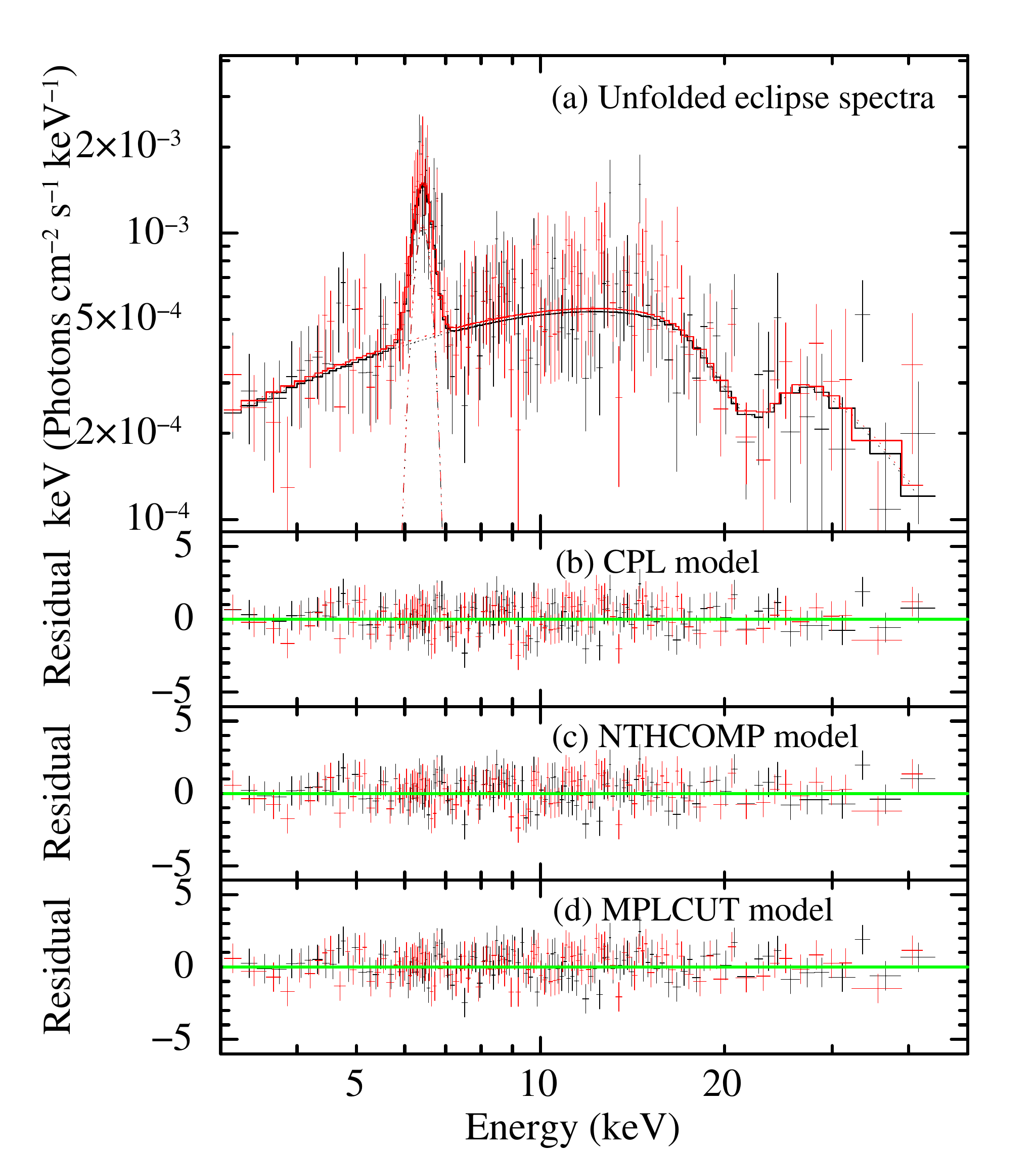}
\caption{Best-fitting FPMA (black) and FPMB (red) eclipse spectra of 4U 1538-522 for the last observation. (a) Unfolded spectra and model components. (b) Residual with CPL model M1 (c) Residual with \texttt{nthcomp} model M2. (d) Residual with \texttt{mplcut} model M3. The dotted and dash-dotted lines represent \texttt{cutoffpl}/\texttt{nthcomp}/\texttt{mplcut} and \texttt{gaus} components, respectively.} 
 \label{fig:spec3b}
 \end{figure}

\begin{figure}
\centering
	\includegraphics[height=\columnwidth,width=\columnwidth]{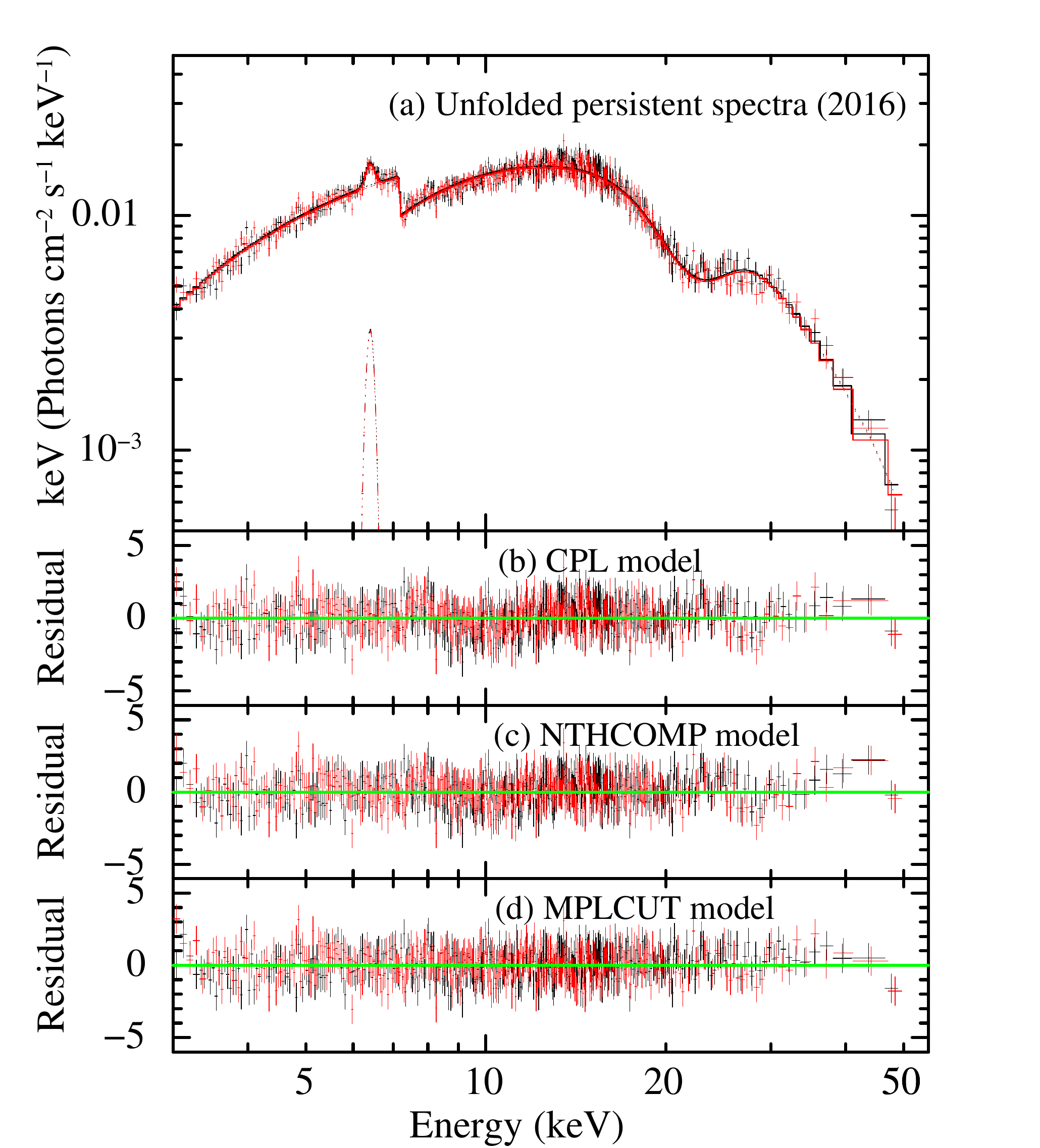}
\caption{Best-fitting FPMA (black) and FPMB (red) persistent spectra for the 2016 August 11 observation. (a) Unfolded spectra and model components. (b) Residual with CPL model M1 (c) Residual with \texttt{nthcomp} model M2. (d) Residual with \texttt{mplcut} model M3. The dotted and dash-dotted lines represent \texttt{cutoffpl}/\texttt{nthcomp}/\texttt{mplcut} and \texttt{gaus} components, respectively.} 
 \label{fig:spec0}
 \end{figure}
 
\begin{figure}
\centering
	\includegraphics[height=\columnwidth,width=\columnwidth]{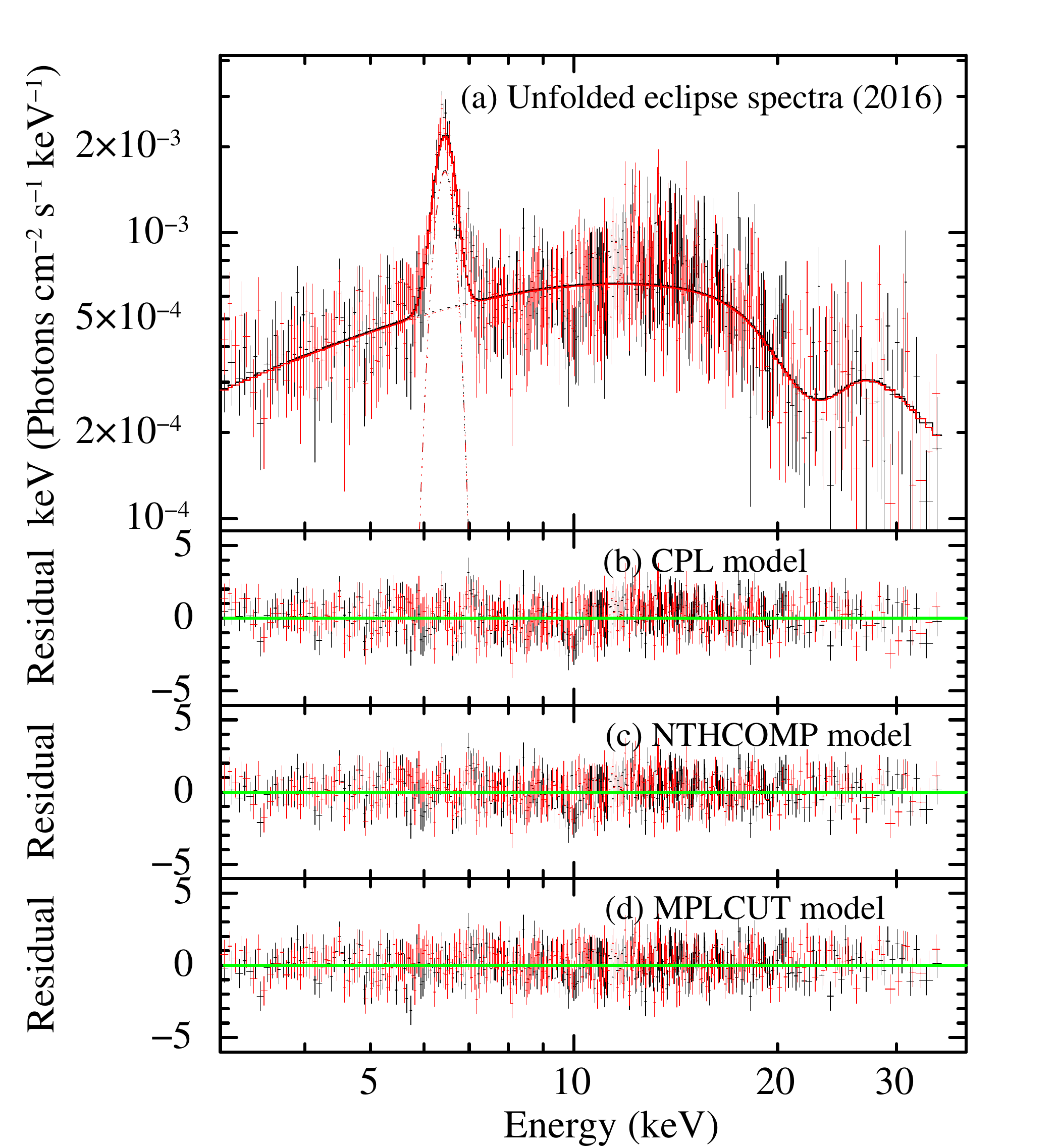}
\caption{Best-fitting FPMA (black) and FPMB (red) eclipse spectra for the 2016 August 11 observation. (a) Unfolded spectra and model components. (b) Residual with CPL model M1 (c) Residual with \texttt{nthcomp} model M2. (d) Residual with \texttt{mplcut} model M3. The dotted and dash-dotted lines represent \texttt{cutoffpl}/\texttt{nthcomp}/\texttt{mplcut} and \texttt{gaus} components, respectively.} 
 \label{fig:ecp0}
 \end{figure}

We then tried the Comptonization models \texttt{compST} \citep{Sunyaev1980} and \texttt{nthcomp} \citep{Zdziarski1996} to model the spectra for a more comprehensive understanding. Similar to the model M1, we included a thermal blackbody component (\texttt{bbodyrad}), a Gaussian emission component (\texttt{gaus}), and an absorbed Gaussian (\texttt{gabs}) for the CRSF to the models. Both the models provided statistically similar fittings with $\chi^2/\nu$ of 1767/1523 and 1777/1523, respectively, along with excess residual near 30 keV and the high-energy end. Addition of a second \texttt{gabs} component improved the fitting significantly with $\Delta \chi^2$ of 141 and 151 for 3 $\Delta$ dof for \texttt{compST} and \texttt{nthcomp} models, respectively. While both the models provide statistically similar fitting, we only present the results for \texttt{nthcomp} model here. We define \texttt{gabs*gabs*(bbodyrad + nthcomp[BB] + gaus)} as model M2. The best-fitting parameters with M2 are reported in the Table~\ref{tab:spec1}.

We repeated the similar exercise for the spectra from the other two observations. The two models provided equally good fitting to the spectra of second observation with the second absorption feature present at energy of $25.42_{-1.33}^{+1.88}$, and $25.73_{-5.18}^{+2.02}$ keV for M1 and M2, respectively. The spectral parameters were consistent with the first observation for both the models. The best-fitting spectra for the second observation are shown in Figure~\ref{fig:spec2} and best-fitting parameters reported in Table~\ref{tab:spec2}.

The third observation covers the late orbital phase (0.803-0.948) and includes the beginning of the eclipse. We extracted the spectra for the pre-eclipse and eclipse region separately. Due to the proximity to the eclipse, the two models did not provide good-fitting to the pre-eclipse spectra with systematic residuals present at the lower energy end. Following the previous works by \citet{Robba2001} and \citet{Hemphill2014,Hemphill2019}, where the low-intensity and near-eclipse spectra are fit by using the partial covering absorption component, we included \texttt{pcfabs} to our models. Addition of this component improved the fitting significantly and returned a high covering column density $N_{\rm H_{\rm cov}}$, of $\sim 27 \times 10^{22}$ cm$^{-2}$ along with a covering fraction of about 0.7 for the two models. A second absorption component was also required to model the pre-eclipse spectra. The best-fitting pre-eclipse spectra of third observation is shown in Figure~\ref{fig:spec3a} and the corresponding parameters are enlisted in Table~\ref{tab:spec3a}.

\begin{table*}
\caption{Best-fitting spectral parameters of 4U 1538-522 for the second observation. All the errors are quoted at 90 per cent confidence level.}
\label{tab:spec2}
\begin{tabular*}{1.8\columnwidth}{c c| c c | c c| c c } 

\hline
& & & & & & &\\ 
\textbf{Component} & \textbf{Parameters} & \multicolumn{2}{c}{\textbf{Model M1}} & \multicolumn{2}{c}{\textbf{Model M2}} & \multicolumn{2}{c}{\textbf{Model M3}} \\
\hline
 & & & & & & &\\

TBABS & $N_{\rm H}$ ($10^{22}$ cm$^{-2}$) &\multicolumn{2}{c}{$-$} &\multicolumn{2}{c}{$ - $} & \multicolumn{2}{c}{$1.1 \pm 0.4$}   \\[0.7ex]

EDGE & $E_{\rm Fe \ \textsc{i} }$ (keV) & $-$ & $-$& $-$&$-$ & $ 7.66_{-0.32}^{+0.16}$ & $7.68_{-0.33}^{+0.15}$ \\[0.5ex]
 & $ \tau$ &$-$ & $-$& $-$& $-$& $ 0.08\pm 0.02$ & $0.08\pm 0.02$ \\[0.7ex]

 BBODYRAD & $kT_{\rm BB}$ (keV)  & $1.15 \pm 0.05 $  & $ 1.16_{-0.04}^{+0.05} $ & $1.18_{-0.06}^{+0.03} $  & $ 1.17_{-0.07}^{+0.03} $ & $-$ & $-$ \\[0.5ex]
 & $N_{\rm BB}$ & $8.16_{-0.78}^{+0.94}$ & $ 7.79_{-0.68}^{+0.82}$  & $6.89_{-0.39}^{+0.69}$ & $ 6.33_{-0.36}^{+0.59}$ & $-$ & $-$ \\[0.5ex]
& ${R_{\rm BB}^a}$ (km) &  $1.8 \pm 0.1$  & $ 1.8\pm 0.1$  & $1.7 \pm 0.1 $ & $ 1.6\pm 0.1 $ & $ - $ & $-$ \\[0.5ex]
& $f^{a}_{\rm bol}$ & \multicolumn{2}{c}{$ 1.50\pm0.01 $} & \multicolumn{2}{c}{$1.28 \pm 0.01$} & \multicolumn{2}{c}{$-$} \\[0.7ex]

 CPL  & $\Gamma$ & $-1.48_{-0.19}^{+0.16} $ & $ -1.31_{-0.19}^{+0.15}$ & $1.11_{-0.04}^{+0.05} $ & $ 1.16_{-0.04}^{+0.07}$ & $1.13 \pm 0.02 $ & $1.14\pm 0.02 $ \\ [0.5ex]
  /NTHCOMP & $E_{\rm cut}$ (keV) & $ 5.02 \pm 0.21 $ & $ 5.33_{-0.23}^{+0.21} $ &$-$ &$-$& $16.15_{-1.51}^{+0.48}$ & $16.52_{-0.53}^{+1.05}$ \\[0.5ex]
 /MPLCUT & $kT_{\rm e}$ (keV) & $-$ & $-$ & $ 4.59_{-0.07}^{+0.10} $ & $ 4.79_{-0.07}^{+0.12} $ & $ - $ & $-$ \\[0.5ex]
 & $E_{\rm fold}$ (keV) &$-$ &$-$ & $-$ & $-$ &$ 10.0 \pm 0.3 $  & $10.1_{-0.68}^{+0.32} $ \\[0.5ex]
  & ${\tau}$  & $-$ & $-$ & $29.8_{-5.8}^{+8.5}$ & $23.7_{-4.6}^{+4.3}$ & $-$ & $-$ \\[0.5ex]

& Norm ($10^{-3}$) & $4.8 \pm 0.2$ & $6.3 \pm 0.2$ & $1.1_{-0.4}^{+0.6}$ & $1.6_{-0.4}^{+0.8}$ & $29.7 \pm 1.4$ & $30.0 \pm 1.3$ \\[0.5ex]
 & $f^{a}_{\rm bol}$ & \multicolumn{2}{c}{$6.90 \pm 0.03 $} & \multicolumn{2}{c}{$ 7.25 \pm 0.03 $} & \multicolumn{2}{c}{$ 9.29\pm 0.03 $} \\[0.7ex]

 GAUS  & $E_{\rm line}$ (keV) & $ < 6.48 $ & $ < 6.47 $ & $< 6.48 $ & $ < 6.46 $ & $< 6.41$ & $ < 6.41 $\\[0.5ex]
 & $\sigma$ (keV) & $ 0.40_{-0.20}^{+0.21} $ & $0.30_{-0.14}^{+0.19}$ & $ 0.31_{-0.14}^{+0.23} $ & $ 0.23_{-0.11}^{+0.18} $ & $0.01^{\dag} $ & $0.01^{\dag}$ \\[0.5ex]

 & Norm ($10^{-4}$) & $4.3_{-1.4}^{+2.0} $ &$3.5_{-1.0}^{+1.5}$ & $3.6_{-1.0}^{+1.9}$ & $3.0_{-0.6}^{+0.8}$ & $1.5 \pm 0.3$ & $1.5 \pm 0.3$ \\[0.5ex]
 & EW (eV) &$126_{-23}^{+28}$ & $101_{-17}^{+22} $ & $104_{-18}^{+20}$ & $87_{-17}^{+18} $ & $42 \pm 8 $ & $42 \pm 8 $\\[0.5ex]

 & $f^{b}_{\rm bol}$  & \multicolumn{2}{c}{$0.36 \pm 0.04 $} & \multicolumn{2}{c}{$0.31 \pm 0.03 $} & \multicolumn{2}{c}{$ 0.16\pm 0.03 $} \\[0.7ex]

 GABS & $E_{\rm line}$ (keV) & $22.24 \pm 0.14 $ & $ 20.85_{-0.41}^{+0.55}$ & $22.29 \pm 0.14 $ & $ 20.96_{-0.38}^{+0.42}$ & $ 21.6 \pm 0.2 $ & $21.38_{-0.88}^{+0.33}$ \\[0.5ex]
 & Width (keV) & $3.32_{-0.16}^{+0.17} $ & $2.50_{-0.30}^{+0.28}$ & $3.44 \pm 0.15  $ & $ 2.61_{-0.34}^{+0.30} $ &  $ 3.13_{-0.46}^{+0.21} $ & $ 2.64_{-0.57}^{+0.31} $\\[0.5ex]
 & Depth & $ 6.67 \pm 0.49 $ & $ 3.60_{-1.33}^{+1.66}$ & $ 7.18_{-0.40}^{+0.42}$ & $ 3.85_{-1.31}^{+1.54}$ & $ 5.15_{-1.38}^{+0.57}$ & $ 4.6_{-2.0}^{+0.6} $ \\[0.7ex]

  GABS & $E_{\rm line}$ (keV) & $-$ & $25.42_{-1.33}^{+1.88}$ & $-$ & $25.73_{-5.18}^{+2.02}$ & $-$ & $26.5_{-2.5}^{+0.7}$ \\[0.5ex]
 & Width (keV) & $-$ &  $ 3.74_{-1.29}^{+0.94}$ & $-$ & $ 4.24_{-1.26}^{+0.93}$ & $-$ & $1.42_{-0.59}^{+1.49} $\\[0.5ex]
 & Depth & $-$ & $3.71_{-1.84}^{+1.55}$ & $-$ & $ 4.40_{-1.97}^{+1.45} $ & $-$ & $ 0.64_{-0.34}^{+1.36}$ \\[0.5ex]

  & & & & & & & \\
 &$f^{a}_{\rm Total}$ & \multicolumn{2}{c}{$8.43 \pm 0.03$} & \multicolumn{2}{c}{$ 8.57 \pm 0.02 $} & \multicolumn{2}{c}{$ 9.31 \pm 0.03 $} \\[0.7ex]

\hline
 & $\chi^2$/dof & 1448.2/1327 & 1382.8/1324 & 1457.9/1327 & 1385.3/1324 & 1376.9/1324 & 1361.5/1321 \\
\hline

 \multicolumn{8}{l}{Model M1 = \texttt{gabs*gabs*(bbodyrad + cutoffpl + gaus)}} \\
 \multicolumn{8}{l}{Model M2 = \texttt{gabs*gabs*(bbodyrad + nthComp[bb] + gaus)}} \\
 \multicolumn{8}{l}{Model M3 = \texttt{tbabs*edge*gabs*gabs*(powerlaw*highecut + gaus + gaus)}}\\
 \multicolumn{8}{l}{$^{a} f$ is the unabsorbed bolometric flux in 0.1--100.0 keV in $10^{-10}$ erg cm$^{-2}$ s$^{-1}$.}\\ 
\multicolumn{8}{l}{$^{b} f$ is the unabsorbed bolometric flux in 0.1--100.0 keV in $10^{-11}$ erg cm$^{-2}$ s$^{-1}$.}\\
\multicolumn{8}{l}{$^{\dag}$ Parameters fixed at the value for fitting.}
\end{tabular*} 
\end{table*}

\begin{table}
\caption{Best-fitting spectral parameters for the pre-eclipse spectra of 4U 1538-522 from the third observation. All the errors are quoted at 90 per cent confidence level.}
\label{tab:spec3a}
\scalebox{0.8}{
\begin{tabular}{c c|c c c } 

\hline
& & & & \\
\textbf{Component} & \textbf{Parameters} & \textbf{Model M1} & \textbf{Model M2} & \textbf{Model M3} \\
\hline
 & & & & \\

 TBABS & $N_{\rm H}$ ($10^{22}$ cm$^{-2}$) & $-$ & $-$ & $ 9.14 \pm 0.63 $\\[0.5ex]

 PCFABS & $N_{\rm H_{cov}}$ ($10^{22}$ cm$^{-2}$) &  $ 27 \pm 7 $ & $ 26 \pm 7$ & $-$ \\[0.5ex]
 & Cvrfrac & $0.71 \pm 0.06 $ & $ 0.70_{-0.05}^{+0.06} $ & $ - $ \\ [0.7ex]

 BBODYRAD & $kT_{\rm BB}$ (keV) & $ 1.06_{-0.10}^{+0.12} $ & $ 1.09_{-0.23}^{+0.09} $ & $- $ \\[0.5ex]
 & $N_{\rm BB}$ & $ 11.5_{-3.8}^{+6.3}$ & $ 9.09_{-3.23}^{+2.48}$ & $-$ \\[0.5ex]
 
& ${R_{\rm BB}^a}$ (km) &  $2.2 \pm 0.3 $ & $ 1.9 \pm 0.3$ & $ - $  \\[0.5ex]
& $f^{a}_{\rm bol}$ & $1.58 \pm 0.02$ & $1.39 \pm 0.02$ & $- $ \\[0.7ex]
 
 CPL  & $\Gamma$  & $ -1.31_{-0.72}^{+0.30}$ & $ < 1.28 $ & $ 0.89 \pm 0.03$  \\ [0.5ex]
 /NTHCOMP & $E_{\rm cut}$ (keV) & $ 5.28_{-0.41}^{+0.40}$ & $-$ & $ 14.88 \pm 0.35 $ \\[0.5ex]
  /MPLCUT & $kT_{\rm e}$ (keV) & $-$ & $ 4.69_{-0.19}^{+0.22}$ & $-$ \\[0.5ex]
  & $E_{\rm fold}$ (keV) &$-$ &$-$ & $9.51 \pm 0.24 $ \\[0.5ex]
  & ${\tau}$  & $-$ & $ > 8$  & $-$ \\[0.5ex]

  & Norm ($10^{-3}$) & $0.7_{-0.4}^{+0.3}$ & $1.5_{-1.1}^{+3.4}$  & $ 17.6 \pm 1.1$ \\[0.5ex]
  & $f^{b}_{\rm bol}$ & $7.64 \pm0.04 $ & $ 8.07\pm 0.04$ & $ 8.75\pm 0.03$\\[0.7ex]

 GAUS & $E_{\rm line}$ (keV) & $ < 6.42 $ & $< 6.42 $ & $ < 6.41 $ \\[0.5ex]
 & $\sigma$ (keV) & $0.30_{-0.18}^{+0.14}$ & $ 0.30_{-0.17}^{+0.13} $ & $0.01^{\dag}$ \\[0.5ex]
 & Norm ($10^{-4}$) & $2.9_{-1.4}^{+1.5}$ & $3.0 \pm 0.1$ & $1.2 \pm 0.4 $ \\[0.5ex]
 &  EW (eV) & $81_{-37}^{+74}$ & $ < 84 $ & $37\pm 11$\\[0.5ex]
 & $f^{b}_{\rm bol}$ & $ 0.30\pm 0.05$ & $0.31 \pm 0.06 $ & $  0.13\pm 0.03$\\[0.7ex]

 GABS & $E_{\rm line}$ (keV) & $21.84_{-1.22}^{+0.38}$  & $ 21.1 \pm 0.5 $ & $21.83 \pm 0.18 $ \\[0.5ex]
 & Width (keV) & $3.11_{-1.12}^{+0.36}$  & $ 2.55_{-0.61}^{+0.88} $ & $ 3.14_{-0.21}^{+0.24} $\\[0.5ex]
 & Depth & $6.3_{-4.9}^{+1.2}$ & $2.78_{-1.45}^{+2.75}$ & $ 4.93_{-0.45}^{+0.51}$ \\[0.7ex]

 GABS & $E_{\rm line}$ (keV) & $28.3_{-4.4}^{+1.2}$ & $24.4_{-1.0}^{+4.6}$ & $-$ \\[0.5ex]
 & Width (keV) & $2.38_{-1.05}^{+3.32}$ & $ 5.03_{-1.55}^{+1.20}$ & $-$\\[0.5ex]
 & Depth & $1.44_{-0.72}^{+7.04}$ & $6.73_{-4.55}^{+1.89}$ & $-$ \\[0.5ex]

  & & & & \\
 &$f^{a}_{\rm Total}$ &$9.25 \pm0.03 $ & $ 9.50 \pm 0.04$ & $8.77 \pm 0.03$ \\[0.7ex] 

\hline
 & $\chi^2$/dof & 1521.8/1475 & 1524.1/1475 & 1539.1/1477 \\
\hline

 \multicolumn{5}{l}{Model M1 = \texttt{gabs*gabs*pcfabs*(bbodyrad + cutoffpl + gaus)}} \\
 \multicolumn{5}{l}{Model M2 = \texttt{gabs*gabs*pcfabs*(bbodyrad + nthComp[bb] + gaus)}} \\
  \multicolumn{5}{l}{Model M3 = \texttt{tbabs*edge*gabs*(powerlaw*highecut + gaus + gaus)}} \\
  \multicolumn{5}{l}{$^{a} f$ is the unabsorbed bolometric flux in 0.1--100.0 keV in $10^{-10}$ erg cm$^{-2}$ s$^{-1}$.}\\ 
\multicolumn{5}{l}{$^{b} f$ is the unabsorbed bolometric flux in 0.1--100.0 keV in $10^{-11}$ erg cm$^{-2}$ s$^{-1}$.}\\
\multicolumn{5}{l}{$^{\dag}$ Parameters fixed at the value for fitting.}
\end{tabular} }
\end{table}

\begin{table}
\caption{Best-fitting spectral parameters for the eclipse spectra of 4U 1538-522 from the third observation. All the errors are quoted at 90 per cent confidence level.}
\label{tab:spec3b}
\scalebox{0.85}{
\begin{tabular}{c c|c c c} 
\hline
\textbf{Component} & \textbf{Parameters} & \textbf{Model M1} & \textbf{Model M2} & \textbf{Model M3} \\
\hline
 TBABS & $N_{\rm H}$ ($10^{22}$ cm$^{-2}$) & $-$ & $-$ & $< 10$\\[0.7ex]

  CPL & $\Gamma$  & $-0.30_{-0.28}^{+0.25} $ & $ 1.22_{-0.07}^{+0.24}$  & $0.26_{-0.11}^{+0.39}$\\[0.5ex]
 /NTHCOMP & $E_{\rm cut}$ (keV) & $ 9.48_{-1.87}^{+2.54} $ & $ - $ & $ 9.41_{-0.16}^{+0.19} $\\[0.5ex]
 /MPLCUT & $kT_{\rm e}$ (keV) & $-$ & $ 5.89_{-0.73}^{+1.52} $ & $ - $\\[0.5ex]
  & $E_{\rm fold}$ (keV) &$-$ & $-$ &$ 12.5_{-2.3}^{+5.2}$ \\[0.5ex]
  & ${\tau}$  & $-$ & $17.7_{-7.7}^{+5.9}$  & $-$ \\[0.7ex]
  & Norm ($10^{-4}$) & $0.74_{-0.24}^{+0.33} $  & $ 0.60_{-0.26}^{+2.61}$ & $ 1.1_{-0.3}^{+1.1} $\\[0.7ex]
  & $f^{a}_{\rm bol}$ & $ 2.46 \pm 0.10 $ & $2.42 \pm 0.09 $ & $2.54 \pm 0.09$ \\[0.7ex]

 GAUS & $E_{\rm line}$ (keV) & $< 6.47 $ & $ < 6.47 $ & $6.41 \pm 0.05 $ \\[0.5ex]
 & $\sigma$ (keV) & $ 0.22 \pm 0.07 $ & $0.23 \pm 0.07 $ & $0.22 \pm 0.08 $ \\[0.5ex]
 & Norm ($10^{-4}$) & $0.89 \pm 0.15 $ &$0.91 \pm 0.17 $ & $0.88 \pm 0.15 $ \\[0.5ex]
 &  EW (eV) &$1348_{-219}^{+382}$ & $ < 1422 $ & $ < 1336 $ \\[0.7ex]
 & $f^{b}_{\rm bol}$ & $ 0.91 \pm 0.13 $ & $ 0.94 \pm 0.13 $ & $ 0.91 \pm 0.13 $\\[0.7ex]

 GABS & $E_{\rm line}$ (keV) & $21.88_{-1.44}^{+2.05} $ & $ 21.8_{-1.4}^{+2.0}$ & $ 21.8_{-1.5}^{+2.5} $ \\[0.5ex]
 & Width (keV) & $2.58_{-1.36}^{+1.63} $ & $2.92_{-1.18}^{+1.83} $ &  $ 2.4_{-1.9}^{+1.8} $\\[0.5ex]
 & Depth & $ 3.78_{-2.07}^{+3.22}$ & $ 4.86_{-2.51}^{+4.35}$ & $ 3.2_{-2.2}^{+2.9}$ \\[0.5ex]

 & & & & \\
 &$f^{a}_{\rm Total}$ &$ 2.56 \pm 0.09 $ & $ 2.51 \pm 0.09 $ & $ 2.63 \pm 0.09 $ \\[0.7ex] 

\hline
 & $\chi^2$/dof & 209.1/226 & 208.1/225 & 201.1/222 \\
\hline
 \multicolumn{5}{l}{\textit{Notes.} Model M1 = \texttt{gabs*(cutoffpl + gaus)}} \\
 \multicolumn{5}{l}{Model M2 = \texttt{gabs*(nthComp[BB] + gaus)}} \\
 \multicolumn{5}{l}{Model M3 = \texttt{tbabs*gabs*(powerlaw*highecut + gaus + gaus)}} \\
 \multicolumn{5}{l}{$^{a} f$ is the unabsorbed bolometric flux in 0.1--100.0 keV in $10^{-11}$ erg cm$^{-2}$ s$^{-1}$.}\\ 
 \multicolumn{5}{l}{$^{b} f$ is the unabsorbed bolometric flux in 0.1--100.0 keV in $10^{-12}$ erg cm$^{-2}$ s$^{-1}$.}
  
\end{tabular} }
\end{table}

\begin{table*}
\caption{Best-fitting spectral parameters for the persistent and eclipse spectra of 4U 1538-522 for 2016 August 11 observation. All the errors are quoted at 90 per cent confidence level.}
\label{tab:spec0}
\begin{tabular*}{1.8\columnwidth}{c c | c c c| c c c } 

\hline
& & & & & & & \\
\textbf{Component} & \textbf{Parameters} & \multicolumn{3}{c}{\textbf{Persistent}} & \multicolumn{3}{c}{\textbf{Eclipse}} \\
& & \textbf{Model M1} & \textbf{Model M2} & \textbf{Model M3} & \textbf{Model M1} & \textbf{Model M2} & \textbf{Model M3}\\
\hline
 & & & & & & & \\
TBABS & $N_{\rm H}$ ($10^{22}$ cm$^{-2}$) & $5 \pm 1$ & $12 \pm 1$ & $12 \pm 1$ & $-$ & $-$ & $7 \pm 5$ \\[0.7ex]

EDGE & $E_{\rm Fe \textsc{i}}$ (keV) & $7.18 \pm 0.03$ & $7.20 \pm 0.06 $& $7.15_{-0.15}^{+0.09}$ & $-$ & $-$ & $-$ \\[0.5ex]
 & $ \tau$ &$0.36 \pm 0.02 $ & $0.22 \pm 0.02 $ & $0.14 \pm 0.02$ & $-$& $-$ & $-$ \\[0.7ex]

PCFABS & $N_{\rm H_{cov}}$ ($10^{22}$ cm$^{-2}$) &  $-$ & $-$ & $-$ & $-$ & $-$ & $131 \pm 17$ \\[0.5ex]
 & Cvrfrac & $-$ & $-$ & $-$ & $-$ & $-$ & $0.92_{-0.03}^{+0.02} $ \\ [0.7ex]

  CPL & $\Gamma$ & $ -0.78\pm 0.06$ & $ 1.24 \pm 0.01$ & $0.70_{-0.05}^{+0.04} $ & $ -0.41_{-0.14}^{+0.13}$ & $1.18_{-0.02}^{+0.03}$ & $ 2.3 \pm  0.3$ \\ [0.5ex]
  /NTHCOMP & $E_{\rm cut}$ (keV) & $ 6.28 \pm 0.14 $ & $ - $ & $ 16.1_{-0.77}^{+0.45} $ & $8.28_{-0.84}^{+1.04} $ & $-$ & $9.97_{-0.38}^{+0.27} $ \\[0.5ex]
  /MPLCUT & $kT_{\rm e}$ (keV) & $-$ & $5.0 \pm 0.1$ & $-$ & $-$ & $5.2 \pm 0.3$ & $-$ \\[0.5ex]
  & $E_{\rm fold}$ (keV) &$-$ &$-$ & $9.12_{-0.38}^{+0.41}$ & $-$ &$-$  & $35_{-14}^{+53} $ \\[0.5ex]
  & ${\tau}$ & $-$ &$ 18.4\pm 0.6 $ & $-$ & $-$ & $ 21.2_{-2.4}^{+2.2} $ & $-$ \\[0.5ex]
 & Norm ($10^{-3}$) & $1.43_{-0.14}^{+0.16} $ & $11.1_{-3.9}^{+2.9}$ & $8.4_{-1.0}^{+0.9}$ & $0.09_{-0.01}^{+0.02}$ & $0.15_{-0.07}^{+0.2} $ & $40_{-24}^{+53}$ \\[0.5ex]
 & $f^{a}_{\rm bol}$ & $6.32 \pm 0.02$ & $6.98 \pm 0.03$ & $7.02 \pm 0.03$ & $ 0.28 \pm 0.01 $ & $0.27 \pm 0.01$ & $0.52 \pm 0.01$ \\[0.7ex]

 GAUS & $E_{\rm line}$ (keV) & $ 6.41 \pm 0.05 $ & $ < 6.48$ & $< 6.52 $ & $ 6.44 \pm 0.02 $ & $6.44 \pm 0.02$ & $6.41 \pm 0.01$\\[0.5ex]
 & $\sigma$ (keV) & $ < 0.19$  & $ 0.19_{-0.08}^{+0.11} $ & $ 0.28_{-0.09}^{+0.18} $ & $0.22 \pm 0.03 $ & $0.24_{-0.03}^{+0.02}$ & $ 0.01^{f} $ \\[0.5ex]
 & Norm ($10^{-4}$) & $1.5 \pm 0.4$ & $2.3_{-0.3}^{+0.7}$ & $3.2_{-0.7}^{+1.3}$ & $1.4 \pm 0.1$ & $1.5 \pm 0.1$ & $ 3.14 \pm 0.07 $ \\[0.5ex]
 & EW (eV) & $65 \pm 19$ & $97_{-27}^{+2} $ & $140_{-31}^{+32}$ & $1697_{-119}^{+152}$ & $< 1900$ & $618_{-65}^{+546}$\\[0.7ex]
 & $f^{b}_{\rm bol}$ & $0.15 \pm 0.03 $ & $0.24 \pm 0.03 $ & $0.33 \pm 0.04$ & $0.15 \pm 0.01 $ & $0.16 \pm 0.01$ & $0.09 \pm 0.01$ \\[0.7ex]

 GABS & $E_{\rm line}$ (keV) & $ 22.14 \pm 0.17$ & $ 22.28_{-0.17}^{+0.18}$ & $21.57 \pm 0.20 $ & $ 22.3_{-0.86}^{+1.09}$ & $21.9_{-0.8}^{+0.9}$ & $ 21.5_{-0.9}^{+1.1}$ \\[0.5ex]
 & Width (keV) & $2.95 \pm 0.16$ & $3.26_{-0.15}^{+0.17}$ & $3.47_{-0.35}^{+0.33} $ & $2.60_{-0.70}^{+0.85}$ &  $ 2.86_{-0.69}^{+0.81} $ & $2.09_{-0.80}^{+1.50}$\\[0.5ex]
 & Strength & $ 4.63_{-0.32}^{+0.33}$ & $ 5.66_{-0.18}^{+0.39}$  & $ 5.56_{-0.95}^{+0.88}$ & $ 3.5_{-1.2}^{+1.7}$ & $ 4.21_{-1.38}^{+1.89}$ & $ 1.89_{-0.88}^{+1.49}$ \\[0.5ex]


 & & & & & & &\\
 &$f^{a}_{\rm Total}$ & $ 6.33 \pm 0.02 $ & $ 7.01 \pm 0.03 $ & $ 7.05  \pm 0.03 $ & $ 0.29 \pm 0.01  $ & $ 0.29 \pm 0.01  $ & $ 0.53 \pm 0.01 $ \\[0.7ex] 

\hline
 & $\chi^2$/dof & 1266.4/1178 & 1252.3/1180 & 1197.8/1175 & 715.4/698 & 694.2/697 &  645.6/693 \\
\hline

\multicolumn{8}{l}{Model M1 = \texttt{tbabs*edge*gabs*(cutoffpl + gaus)}} \\
\multicolumn{8}{l}{Model M2 = \texttt{tbabs*edge*gabs*(nthComp[bb] + gaus)}} \\
\multicolumn{8}{l}{Model M3 = \texttt{tbabs*edge*gabs*(powerlaw*highecut + gaus + gaus)}}\\ 
\multicolumn{8}{l}{$^{a} f$ is the unabsorbed bolometric flux in 0.1--100.0 keV in $10^{-10}$ erg cm$^{-2}$ s$^{-1}$.}\\ 
\multicolumn{8}{l}{$^{b} f$ is the unabsorbed bolometric flux in 0.1--100.0 keV in $10^{-11}$ erg cm$^{-2}$ s$^{-1}$.}
\end{tabular*} 
\end{table*}

\begin{figure}
\centering
	\includegraphics[height=0.8\columnwidth]{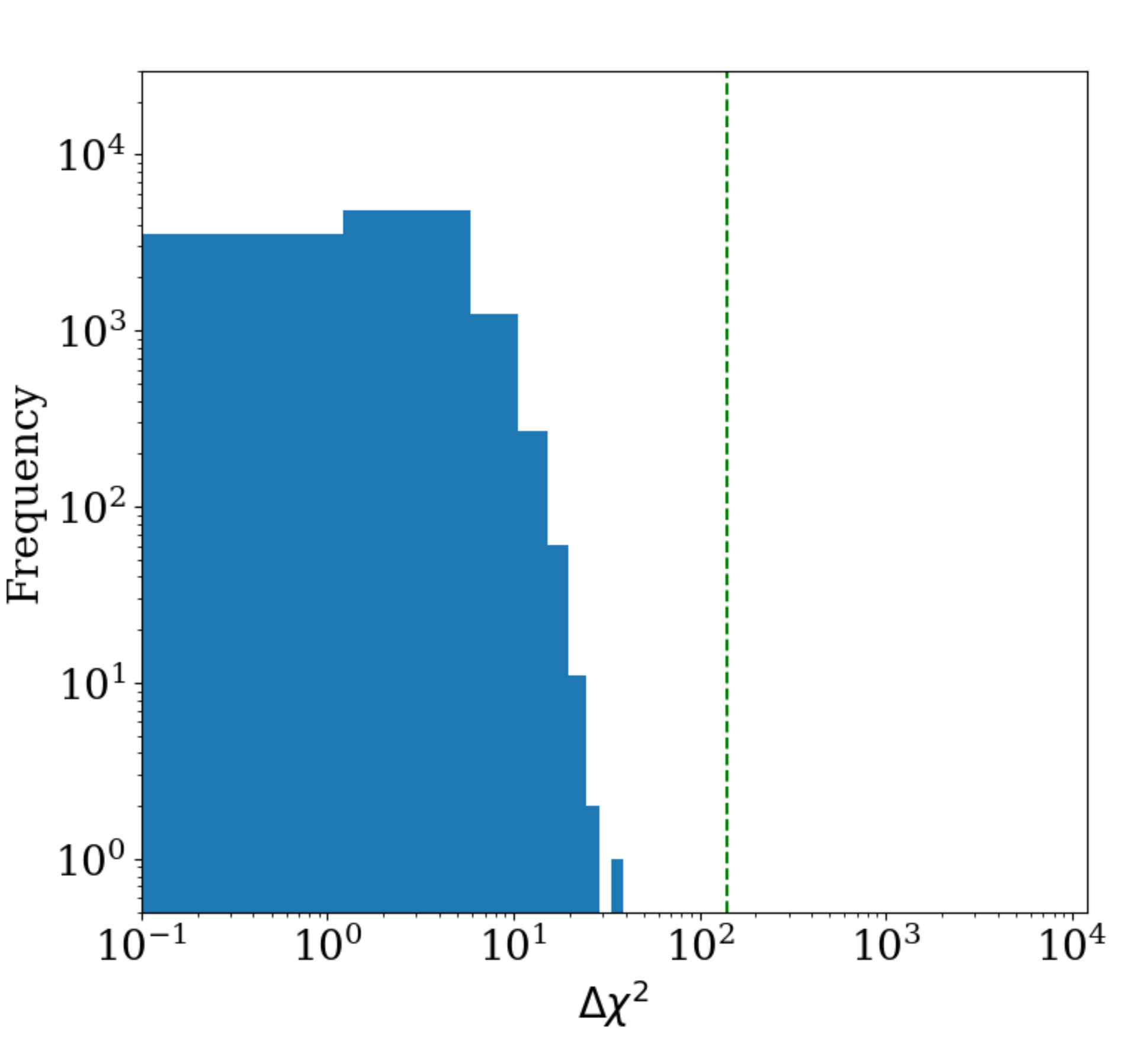}
 	\includegraphics[height=0.8\columnwidth]{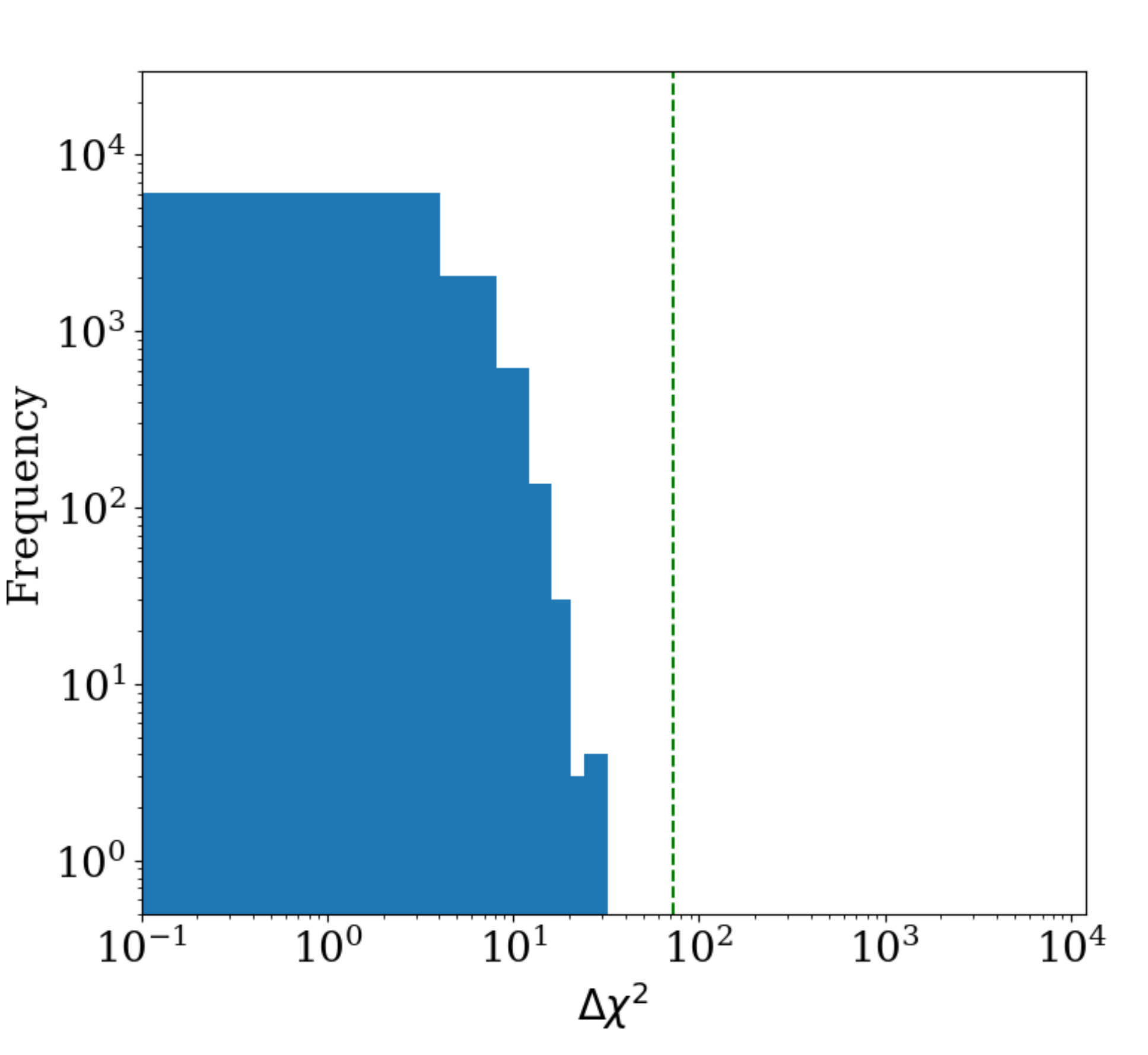}
\caption{Histograms from the $10^{4}$ \texttt{simftest} simulations with model M2, to test the statistical significance of the second absorption feature in the spectra. The top and bottom plots correspond to the simulations for the first and second observations, respectively. The dashed vertical line mark the observed $\Delta \chi^2$ in each plot.} 
 \label{fig:simf}
 \end{figure}

We tried to fit the eclipse spectra separately by using the CPL and Comptonization models. The best-fitting did not favour the blackbody component. We found a weak CRSF present around 22 keV. However, the second absorption feature was not present in the eclipse spectra. Thus, the final models comprised a \texttt{gabs} component for CRSF, a \texttt{gaus} component for Fe emission line and the CPL or Comptonization component (Fig~\ref{fig:spec3b}). The equivalent width (EW) of the emission line was very high along with a slightly higher electron temperature for the Comptonization medium ($kT_{\rm e}$) of $\sim 6$ keV as compared to the results from other spectra (Table~\ref{tab:spec3b}). 

For a comparative study, we extended the spectral analysis to include the \emph{NuSTAR}'s 2016 observation (Obs\_ID 30201028002) which has been already studied by \citet{Hemphill2019}. We used the standard reduction and calibration process as done for other observations of current work. We resolved the entire spectrum into the persistent emission spectrum, including the segments 0--4, 13, and 14 of \citet{Hemphill2019} and eclipse spectrum, comprising segments 5--12. We extracted the source and background spectra by using circular regions of 60 arcsec and re-binned the spectra to contain 25 and 10 counts per energy bin, respectively. We performed the spectral analysis in the energy ranges 3--50 and 3-35 keV for the persistent and eclipse spectra, respectively.

We used our models M1 and M2 to fit the persistent and eclipse spectra. The models provided reasonably good fitting to the persistent spectrum after inclusion of \texttt{tbabs} component, along with an absorption edge at 7 keV. However, the blackbody and the second absorption components were not statistically favoured hence we removed these components from the models. Both the models provided fairly consistent results. The best-fitting spectra are shown in Figure~\ref{fig:spec0} and Table~\ref{tab:spec0} gives the best-fitting parameters. We used the modified models M1 and M2 to model the eclipse spectrum of this observation. Both the models provided equally good fitting with consistent spectral parameters (Table~\ref{tab:spec0}). We show the best-fitting eclipse spectra of this observation in Figure~\ref{fig:ecp0}.

We then used the modified cut-off power law model, \texttt{mplcut}, as used by \citet{Hemphill2019} in order to compare the spectral results from all the four observations during persistent emission. We started with the simple \texttt{mplcut} model with a \texttt{tbabs} component, Gaussian emission lines at 6.4 keV and 7 keV, \texttt{gabs} component for CRSF at 22 keV, and its first harmonic at 50 keV. But, the 7 keV and CRSF harmonic components were not statistically favoured in the fitting model. We therefore, removed these components from the model. Instead, we included an absorption edge at 7 keV. The addition of this component improved the fitting and resolved the excess residual. Thus, we define our \texttt{mplcut} model as \texttt{tbabs*edge*gabs*(powerlaw*highecut + gaus$_{\rm smooth}$ + gaus)} (M3, hereafter). The best-fitting spectral parameters are consistent with the result of \citet{Hemphill2019} (Table~\ref{tab:spec0}). For the eclipse spectrum, model M3 provided a statistically good fitting with a \texttt{pcfabs} component and without the edge component for a $\chi^2 / \nu$ of 645.6/694. 

The model M3 provided reasonably good fitting to all the other spectra. The lower limit for Fe emission line energy was fixed at 6.4 keV. The line width could not be constrained. Therefore, we fixed the line width at 0.01 keV for spectral fitting. The fitting returned a low column density of $\sim 10^{22}$ cm$^{-2}$ for the first and second spectra. We also observed the presence of a weak absorption like feature around 27 keV in the residuals for first and second spectra. We included a second \texttt{gabs} component to model this feature. The inclusion of this component did not provide a statistically significant improvement to the fitting ($\Delta \chi^2$ of 25 and 15 for 3 additional dof, respectively) as compared to the models M1 and M2. The second absorption feature was not statistically favoured for the pre-eclipse spectra of third observation. For the eclipse spectra of third observation, model M3 provided an acceptable fitting with a $\chi^2/\nu$ of 201.1/222, with only one CRSF at $21.8_{-1.5}^{+2.5}$ keV. The other spectral parameters were consistent with the previous models. The best-fitting spectral parameters with model M3 for all three observations are given in respective tables.
 
The second absorption feature was detected in the residual from all three spectra with all the models except pre-eclipse spectrum of third observation with M3. The inclusion of another \texttt{gabs} component to model this feature provided significant improvement to each of the three spectra with all models. In order to further check the statistical significance of this feature in the spectra, we used the \texttt{simftest} routine of the \textsc{xspec}. This routine runs Monte Carlo simulations to generate simulated spectra based on the real observed spectra and then evaluates the difference in $\chi^2$ for any additional model component (\texttt{gabs} in our case). We simulated $10^4$ spectra, each for observations 1 and 2 and used the \texttt{nthcomp} models for the $\Delta \chi^2$ evaluation for the addition of the second absorption component. Histograms plotted in Figure~\ref{fig:simf} shows the results of \texttt{simftest}. The large difference between the observed $\Delta \chi^2$ and maximum value estimated from simulations of $\sim$ 90 and 43, for observations 1 and 2, respectively, implies $> 3\ \sigma$ confidence level significance detection of the second feature in the spectra.

\section{Discussion}
We have studied the timing and broad-band spectral properties of the accretion powered HMXB pulsar, 4U 1538-522 with the \emph{NuSTAR} archival data and \emph{Fermi}/GBM monitoring data. We have used the archived \emph{NuSTAR} observations, one from May 2019 and two from February 2021 for our analysis. We have also utilized the \emph{Fermi}/GBM monitoring data covering about 14 yr and combined with the previous results from literature, to study the long-term spin trends in 4U 1538-522. Our results present three interesting aspects, 
\begin{enumerate}[leftmargin=0.3cm]
\item Detection of a recent torque reversal in 4U 1538-522 along with sinusoidal variation in the current spin-up phase
\item Comparatively higher spin-up rate after torque reversal with $\dot{P} = -14.4(3) \times 10^{-9}$ s s$^{-1}$ compared to $-1.9(1) \times 10^{-9}$ s s$^{-1}$ during the current sinusoidal phase.
\item Presence of two closely spaced absorption features in the emission spectra.
\end{enumerate}

We have found X-ray pulsations up to 60 keV at $526.57(6)$, $526.2(1)$, and $526.3(1)$ s in the three \emph{NuSTAR} observations, respectively. The spin period decreases from first observation to the second and third observations implying a spin-up of the source between this period. This trend is consistent with the \emph{Fermi}/GBM results from the same observation span. 



The average rate of change of spin-period before and after the third torque reversal is almost consistent $\approx 10^{-9}$ s s$^{-1}$ (Table~\ref{tab:pderiv}), same as before for the previous two reversals. Permanent torque reversals have been observed for other accretion-powered pulsars like GX 1+4 \citep{Chakrabarty1997,Gonzalez2012}, Her X-1 \citep{Staubert2006}, Cen X-3 \citep{Finger1994}, Vela X-1 \citep{Tsunemi1989}, and OAO 1657-415 \citep{Sharma2022a}. While \citet{Rubin1997} suggested that the random walk behaviour in period derivative post 1990 could be responsible for the first observed torque reversal of 4U 1538-522, the following spin-up trend lasting for about 20 yr did not favor this proposition. 

As per the standard accretion torque theory, torque and accretion rate should be correlated. But this also fails in case of 4U 1538-522 \citep{Rubin1997,Hemphill2013}. In case of Vela X-1, it has been presumed that long-term changes in the stellar wind properties could be responsible for sudden torque reversals \citep{Hayakawa1982}. It is likely that 4U 1538-522 is experiencing a similar activity in its companion. There are theories in literature about possibility of a transient accretion disc formation, which can efficiently transfer large-angular momentum to the NS on a variable time scale \citep{Taam1989,Hemphill2013}. The hydrodynamic simulations for wind-fed sources show that formation and dissipation of transient accretion discs with changing rotation directions can also impart the alternating spin-period behaviours resulting in torque reversals \citep{Nelson1997,Rubin1997}.


In the more recent \emph{Fermi}/GBM data, we have found a sinusoidal variation in the pulse period after MJD 59110. Enveloping the over-all spin-up, the sinusoid had a periodicity of $271.2 \pm 2.6$ d and an amplitude of $0.059 \pm 0.004$ s. Such periodic variations have not been reported so far in 4U 1538-522. However, similar short-term fluctuations in the spin-period have been found in several X-ray pulsars at different periods \citep{Gonzalez2012,Molkov2017,Deo2021,Sharma2022a}.

These fluctuations are often attributed to either variations in torque due to the accreting material \citep{Elsner1976,Ghosh1979} or changes in the super-fluid core of the NS which triggers internal torque fluctuations \citep{Lamb1978}. These fluctuations have also been attributed to the short-term changes in the accretion rate for sources like Her X-1 \citep{Staubert2006}, GX-1+4 \citep{Gonzalez2012}, Cen X-3 \citep{Tsunemi1989}, and OAO 1657-415 \citep{Sharma2022a}. According to \citet{Nagase1989}, the irregular stellar winds from early-type companions can also lead to such short-term reversals of accretion torque on NS on time-scales of days to years.

Similar to reported works, the pulse profile shows significant evolution with energy for all the three observations \citep{Robba2001,Hemphill2013}. This evolution is similar to that observed for other accreting X-ray pulsars like Vela X-1 \citep{Kreykenbohm2002} and 4U 0115+63 \citep{Ferrigno2009}. The pulse profile did not change significantly before and after the third torque reversal. 


We have performed the broad-band phase-averaged spectral analysis, covering 3--60 keV energy range. The X-ray continuum is well described by a thermal blackbody component and either a cutoff power law or a Comptonization component along with additional emission and absorption features. The spectral parameters were observed to evolve modestly across the three observations with significant changes evident in the pre-eclipse spectra of last observation. The blackbody temperature ($kT_{\rm BB}$) varies between 0.9 and 1.1 with being lowest for the pre-eclipse spectra and disappears during the eclipse. The estimated blackbody emission radius ($\sim 2$ km) is consistent with the emission from the NS surface/boundary layer \citep{Cackett2010,Sharma2020,Sharma2022c}.

The power law parameters does not vary significantly and while $E_{\rm cut}$ is consistent across three observations (excluding eclipse) it is significantly lower as compared to the previous results \citep{Robba2001,Hemphill2016,Maitra2019}. However, the difference in the value is possibly influenced by the different model choices in these works. The results for the Comptonization models show that the broad-band spectra of 4U 1538-522 can also be described by the thermal Comptonization of the blackbody emission by an optically thick and moderately hot electron corona of temperature $kT_{\rm e} \approx 5$ keV. We have also found the Fe K$_{\alpha}$ emission line around 6.4 keV in the spectra from all three observations. The EW of the emission line is significantly high during the eclipse and is consistent with the results from \citet{Hemphill2019}. The increased EW is generally observed during the partial obscuring of the sources \citep{Jaiswal2015,Sharma2022a}.

CRSFs have been regularly reported in the emission spectrum of 4U 1538-522 \citep{Nagase1989,Clark1990,Robba2001,Rodes2009,Hemphill2013,Hemphill2016}. We have also found the CRSF in the spectra of all three observations around 22 keV. Our measured CRSF energies are consistent within errors with the results of \citet{Hemphill2016} and \citet{Maitra2019}. Interestingly, we also found a second, relatively weaker, absorption feature in the spectra of 4U 1538-522 between 25 and 28 keV with $> 3 \sigma$ confidence level significance. The feature was present independent of the model choice in the spectra of all observations except during the eclipse. We modelled this feature by using the \texttt{gabs} component. While the presence of multiple CRSFs is generally observed in the spectra of HMXBs \citep{Ferrigno2011,Maitra2013,Furst2015}, the features are usually found to be harmonically spaced with respect to the fundamental line energy. Anharmonicities
in the line energy ratio have also been reported in sources such as 4U 0115+63 \citep{Heindl1999}, Vela X-1 \citep{Kreykenbohm2002}, and V0332+53 \citep{Pottschmidt2005}. However, the ratio of the two lines is close to 1 for our case and rules out the possibility of the two features to be related. This is the first time that this second absorption features have been detected in the spectra of 4U 1538-522.

Similar closely spaced CRSFs have been found in the spectra of two other HMXBs, GX 301-2 with two CRSF at 35 and 51 keV \citep{Furst2018} and GRO J1750-27 with CRSF at 36 and 42 keV \citep{Sharma2022c}. The different yet close values of the CRSF lines can possibly be accounted for by considering the different optical depths for the emission origin heights above the NS surface, as the magnetic field strength decreases with increasing height \citep{Rodes2009}. Apparently, \citet{Bulik1992,Bulik1995} discussed the possibility of a broader distribution of the magnetic field than expected for a dipolar field for 4U 1538-522. Moreover, the inability of the simple Gaussian absorption profile to model the complex and distorted shape of a single, broad absorption feature can also result in the requirement for a second component \citep{Schonherr2007,Schwarm2017,Sharma2022c}. 

The two CRSFs are found at $21.41_{-0.26}^{+0.45}$ and $27.53_{-2.76}^{+2.01}$ keV for the first observation, at $20.85_{-0.41}^{+0.55}$ and $25.42_{-1.33}^{+1.88}$ keV for the second observation; and at $21.84_{-1.22}^{+0.38}$ and $28.3_{-4.4}^{+1.2}$ keV in the pre-eclipse spectra of third observation (with M1). We estimate average magnetic field strengths of $1.84_{-0.06}^{+0.04} (1+z) \times 10^{12}$ and $2.33_{-0.24}^{+0.15} (1+z) \times 10^{12}$ G corresponding to the two CRSF, respectively. These values are in agreement with the reported magnetic field strength of $1.7 (1+z) \times 10^{12}$ G \citep{Clark1990}, $1.8 (1+z) \times 10^{12}$ G \citep{Robba2001}, and $1.8-1.9 (1+z) \times 10^{12}$ G \citep{Hemphill2013,Hemphill2014}.


Our measured CRSF energies with \texttt{mplcut} and other two models are consistent within errors with the results of \citet{Hemphill2019} and \citet{Maitra2019}. The centroid energy for the fundamental CRSF is close to 22 keV, which confirms the long-term increase of the CRSF energy, as reported by \citet{Hemphill2016}. Such long-term increase in the CRSF energy may occur due to the changes in the magnetic field strength at the resonance scattering region possibly due to the accumulation of the accretion matter on the NS surface \citep{Mukherjee2012}. Further studies covering longer time-scales can help in better understanding the physical origins of such energy evolution.


\section*{Acknowledgements}

We have utilized the archived \emph{NuSTAR} data provided by the High Energy Astrophysics Science Archive Research Center (HEASARC) online service maintained by the Goddard Space Flight Center. For this work we have made use of the \emph{NuSTAR} Data Analysis Software (NuSTARDAS) jointly developed by the ASI Space Science Data Center (SSDC, Italy) and the California Institute of Technology (Caltech, USA). PS acknowledges the financial support from the Council of Scientific \& Industrial Research (CSIR) under the Senior Research Fellowship (SRF) scheme.

\section*{Data Availability}

The NuSTAR data used for this work can be accessed from the HEASARC online service at \url{https://heasarc.gsfc.nasa.gov/cgi-bin/W3Browse/w3browse.pl}. The archived spin period history of 4U 1538-522 with CGRO/BATSE is available at \url{https://gammaray.nsstc.nasa.gov/batse/pulsar/data/archive.html}. The latest Fermi/GBM period data for 4U 1538-522 are available at \url{https://gammaray.nsstc.nasa.gov/gbm/science/pulsars/lightcurves/4u1538.html}.



\bibliographystyle{mnras}
\bibliography{4U1538} 








\bsp	
\label{lastpage}
\end{document}